\documentclass[a4paper,11pt]{article}
\pdfoutput=1 

\usepackage{jinstpub} 

\usepackage{multirow}
\usepackage{graphicx}
\usepackage{chemfig,siunitx}
\usepackage{epstopdf}
\usepackage{amsmath}
\usepackage{mathrsfs}
\usepackage{amsfonts}
\usepackage{amssymb}
\usepackage{textcomp}
\usepackage{subfig}
\usepackage{enumitem}

\usepackage[sort&compress]{natbib}

\newcommand{\Nucl}[2]{$\mathrm{^{#2}#1}$}

\title{MIMAC low energy electron-recoil discrimination measured with fast neutrons} 

\author[a]{Q.~Riffard}
\author[a]{D.~Santos}
\author[a]{O.~Guillaudin}
\author[a]{G.~Bosson}
\author[a]{O.~Bourrion}
\author[a]{J.~Bouvier}
\author[a]{T.~Descombes}
\author[a]{J.-F.~Muraz}
\author[b]{L.~Lebreton}
\author[b]{D.~Maire}
\author[c]{P.~Colas}
\author[c]{I.~Giomataris}
\author[d]{J.~Busto}
\author[d]{D.~Fouchez}
\author[d]{J.~Brunner}
\author[d,e]{C.~Tao}

\affiliation[a]{LPSC, Universit\'e Grenoble-Alpes, CNRS/IN2P3, Grenoble, France}
\affiliation[b]{APC, Université Paris Diderot, CNRS/IN2P3, CEA/Irfu, Obs de Paris, Sorbonne Paris Cité, 75205 Paris, France}
\affiliation[c]{LMDN, IRSN Cadarache, 13115 Saint-Paul-Lez-Durance, France}
\affiliation[d]{IRFU, CEA Saclay, 91191 Gif-sur-Yvette, France}
\affiliation[e]{Aix Marseille Université, CNRS/IN2P3, CPPM UMR 7346, 13288, Marseille, France}
\affiliation[f]{Tsinghua Center for Astrophysics, Tsinghua University, Beijing 100084, China}

\emailAdd{riffard@apc.in2p3.fr}
\emailAdd{santos@lpsc.in2p3.fr}

\abstract{
	MIMAC (MIcro-TPC MAtrix of Chambers) is a directional WIMP Dark Matter detector project. 
	Direct dark matter experiments need a high level of electron/recoil discrimination 
	 to search for nuclear recoils produced by WIMP-nucleus elastic scattering. 
	In this paper, we proposed an original method for electron event rejection based on a multivariate analysis
	applied to experimental data acquired using monochromatic neutron fields. 
	This analysis shows that a $10^{5}$ rejection power is reachable for electron/recoil discrimination.
	Moreover, the efficiency was estimated by a Monte-Carlo simulation showing that a $10^{5}$ electron rejection power is reached with a $86.49 \pm 0.17\%$ nuclear recoil efficiency considering the full energy range and $94.67 \pm 0.19\%$ considering a 5~keV lower threshold.}

\arxivnumber{1602.01738}


\begin{document}


\maketitle 
\flushbottom

\section*{Introduction}

In the standard model of cosmology the Dark Matter (DM) is about six times more abundant than the baryonic component of the matter in the Universe.
Furthermore, an increasing number of astrophysical observations from local to large scales support this hypothesis.
At the local scale, a dense DM halo should surround the Milky Way.
Due to the relative motion of the solar system with respect to the galactic DM halo, a WIMP flux should be detected on earth.
 The WIMP (Weakly Interacting Massive Particle) is a massive DM particle candidate ($m_{\mathrm{WIMP}} \sim  (1-100 \mathrm{GeV})$) interacting only by weak and gravitational interactions.
 Many other DM particle candidates are proposed but the WIMP is one of the best motivated and able to be explored by direct detection.
The direct detection search strategy goal is the energy spectrum measurement of nuclear recoils produced by WIMP scattering on detector target nuclei in order to constrain the DM particle properties.
Direct detection experiments such as LUX~\cite{Akerib2014}, Xenon~\cite{Aprile2012,Aprile2013}, DakSide~\cite{Agnes2015ftt}, EDELWEISS~\cite{EDELWEISSCollaboration2012}, 
CDMS~\cite{Agnese}, COUPP~\cite{Behnke2011} and KIMS~\cite{Lee2014} put constrains on the WIMP mass,
Spin Independent (SI) and Spin Dependent (SD) cross-sections.
One major limitation of these search strategies arises from the neutron background.
Indeed, this uncharged particle, colliding elastically with the target nuclei, will produce the same searched signal in the detector,
a nuclear recoil with some tenths of~keV of kinetic energy.
The directional detection search strategy, 
first proposed in 1988~\cite{Spergel1988}, is based on the angular distribution of WIMP momentum directions that should present an anisotropy in galactic coordinates. 
Thus, the angular distribution of recoils produced by a scattering of WIMPs on nuclei should present an anisotropy pointing towards the constellation Cygnus.
Ultimate background events, mainly neutrons, should follow an isotropic distribution in galactic coordinates.
Using a profile likelihood analysis~\cite{Billard2012} it has been shown that it is possible to extract a DM signal from background events. 
Moreover, this detection strategy can be used to constrain the DM particle and the halo properties as shown in~\cite{Billard2010}.

As other directional detection experiments~\cite{Ahlen2009},
the aim of the MIMAC project is the measurement of the nuclear recoil energies and their angular distribution to search for this signature.
In order to reach this objective and before applying the directionality degree of freedom a performant electron/recoil discrimination is required.
In a previous work~\cite{Billard2012b}, a boosted decision trees (BDT) analysis was applied on simulations to define the MIMAC low energy electron-recoil discrimination.
In this paper, we propose to determine the electron/recoil discrimination from experimental data acquired with a monochromatic neutron field of 565 keV mean energy.

\section{Experimental set-up and neutron data-taking}

\begin{figure}[tbp]
\centering 
		\includegraphics[width=0.8\linewidth]{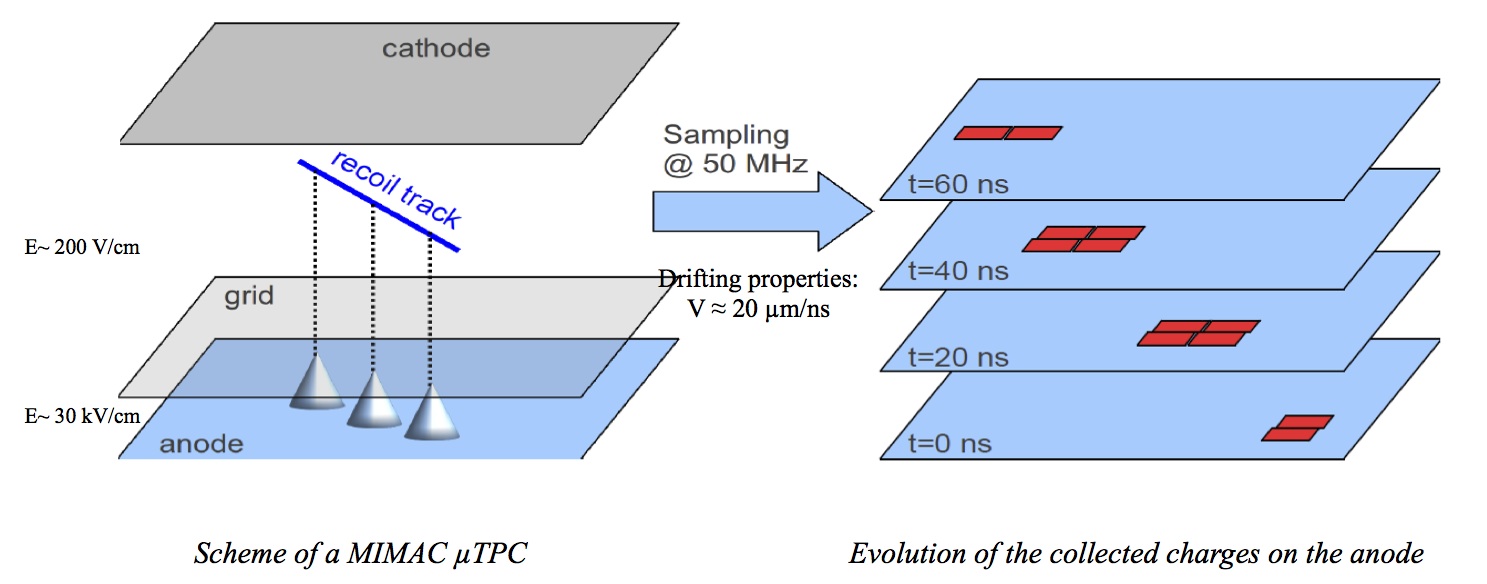}
		\caption{ MIMAC detection strategy principle. The primary ionization electrons are collected to the grid and then amplified in the micromegas gap. The pixel sampling at 50~MHz allows a 3D track reconstruction. }
		\label{fig:Schema_Principe}
\end{figure}

\subsection{The MIMAC detector}

The MIMAC detector~\cite{Santos2007} is a~\textmu-TPC matrix of chambers filled with a low pressure (50 mbar) 
$\mathrm{CF_4} + 28\% \mathrm{CHF_3} + 2\% \mathrm{C_4H_{10}} $ gas mixture.
 The main purpose of this detector is the measurement of nuclear recoil 3D tracks and the estimation of their kinetic energies.

As schematically illustrated in figure~\ref{fig:Schema_Principe},
when a charged particle or a nuclear recoil moves throughout the gas it releases part of its energy by ionization creating electron-ion pairs.
These primary ionization electrons are collected, by an electric field ($E_{drift} =180\,\mathrm{V.cm^{-1}}$),
to the grid of a pixelated bulk micromegas~\cite{Iguaz2011, Giomataris2006} of 10.8~cm side. 
The 200~\textmu m pixels are linked by strips with a 424~\textmu m pitch. 
Passing through the grid, the primary ionization electrons are amplified by avalanche in the 256~\textmu m gap
by a much higher electric field ($E_{gain} =18.36\,\mathrm{kV.cm^{-1}}$).
The pixelated micromegas is coupled to a fast self-triggered electronics (sampled at 50 MHz) specially developed for the MIMAC detector~\cite{Richer2009,Bourrion2010a}.
The read-out composed of 512 channels, 256 channels covering the X-axis and 256 the Y-axis,
allows the measurement of the ionization energy and the description of the envelope of the tracks of nuclear recoils with kinetic energies down to a few~keV depending on the gas and pressure~\cite{Santos2007}. 
Each channel out of 512 has its own threshold determined by a calibration algorithm defining the intrinsic electronic noise level for each channel.
In addition, the total ionization energy is measured by a charge preamplifier connected to the grid coupled to a flash-ADC sampled at the same frequency as the strip channels (50 MHz).

In order to prevent gain degradation due to the presence of impurities and $\mathrm{O_2}$, a closed circuit circulation gas system was implemented.
The circulation system includes a buffer volume, an oxygen filter, a dry and very low leak pump (3.8 $10^{-5}~ mbar. L/s$) and a pressure regulator.
The gas is forced to circulate passing through the oxygen filter renewing the gas in the volume of the chamber every hour.

\begin{figure}[tbp]
\centering 
	\includegraphics[width=0.8\linewidth]{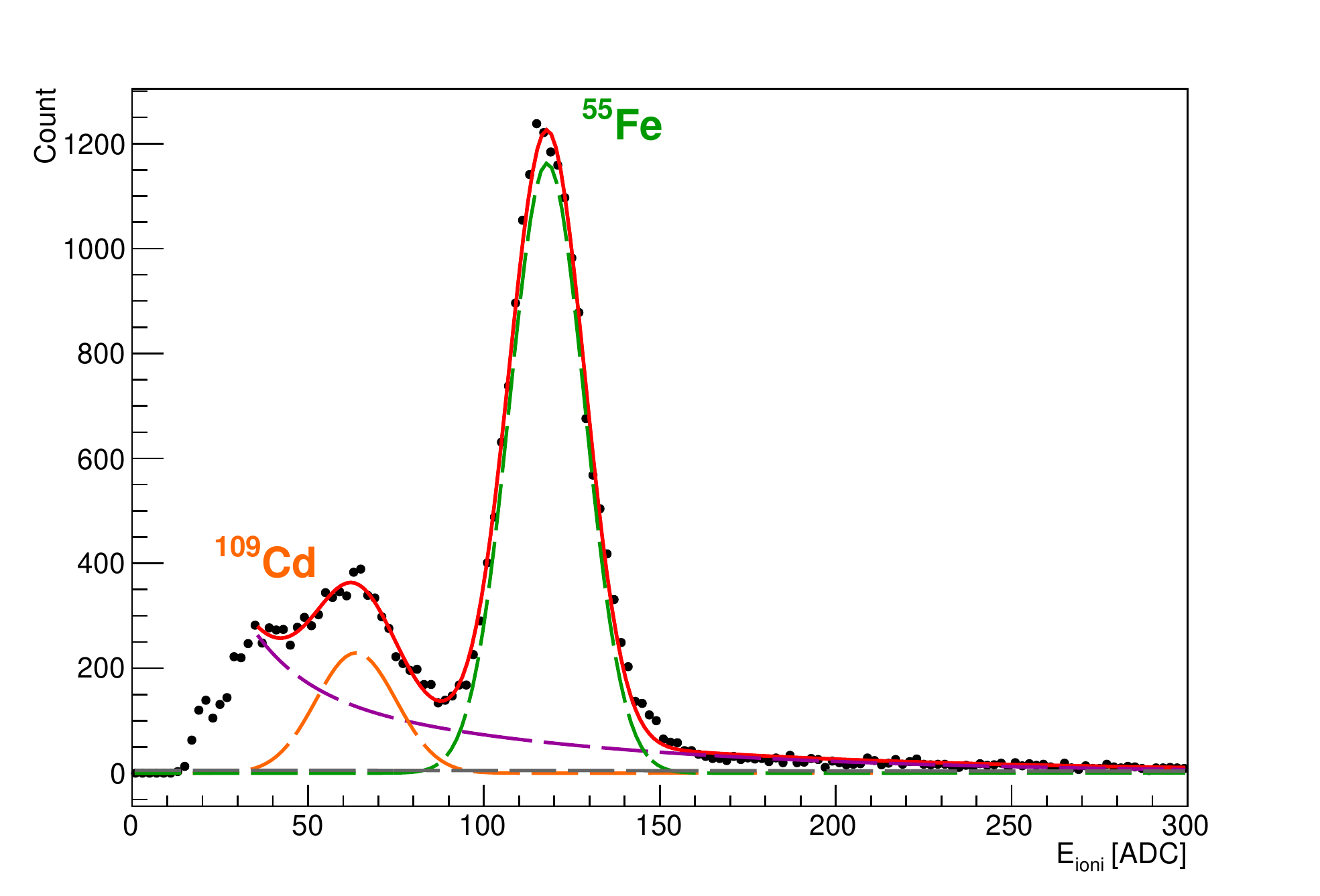}
	\caption{ X-ray calibration spectrum measured with $\mathrm{^{109}Cd}$ and $\mathrm{^{55}Fe}$ sources.
		      $\mathrm{^{109}Cd}$ and $\mathrm{^{55}Fe}$ sources produce X-rays at 3.04 and 5.96~keV respectively. The fit of the total spectrum is shown by the red-solid line showing the peak and background fits by dash lines.
		      }
	\label{fig:Calibration}
\end{figure}

\subsubsection{Ionization energy calibration}
\label{Sec:calib}

The detector calibration was performed by means of two X-rays radioactive sources: 
the $\mathrm{^{109}Cd}$ and $\mathrm{^{55}Fe}$ sources emitting X-rays respectively of 3.04 and 5.96~keV mean energies.
Figure~\ref{fig:Calibration} shows the measured calibration spectrum.
Two peaks can be identified which correspond to $\mathrm{^{109}Cd}$ and $\mathrm{^{55}Fe}$ sources on a continuum background. 
The continuum can be associated with Compton electrons and incomplete charge collection from the 22 keV KX-ray of $\mathrm{^{109}Cd}$. 
This energy spectrum was fitted by the sum of two gaussian functions for the peaks and by the sum of two decreasing exponential functions for the continuum background.
The MIMAC detector shows an energy resolution of 16~\% at 3~keV.
The radioactive sources have been hidden behind separated valves during the neutron detection.

\subsection{MIMAC @ AMANDE facility}

\begin{figure}[tbp]
\centering 
		\includegraphics[width=0.8\linewidth]{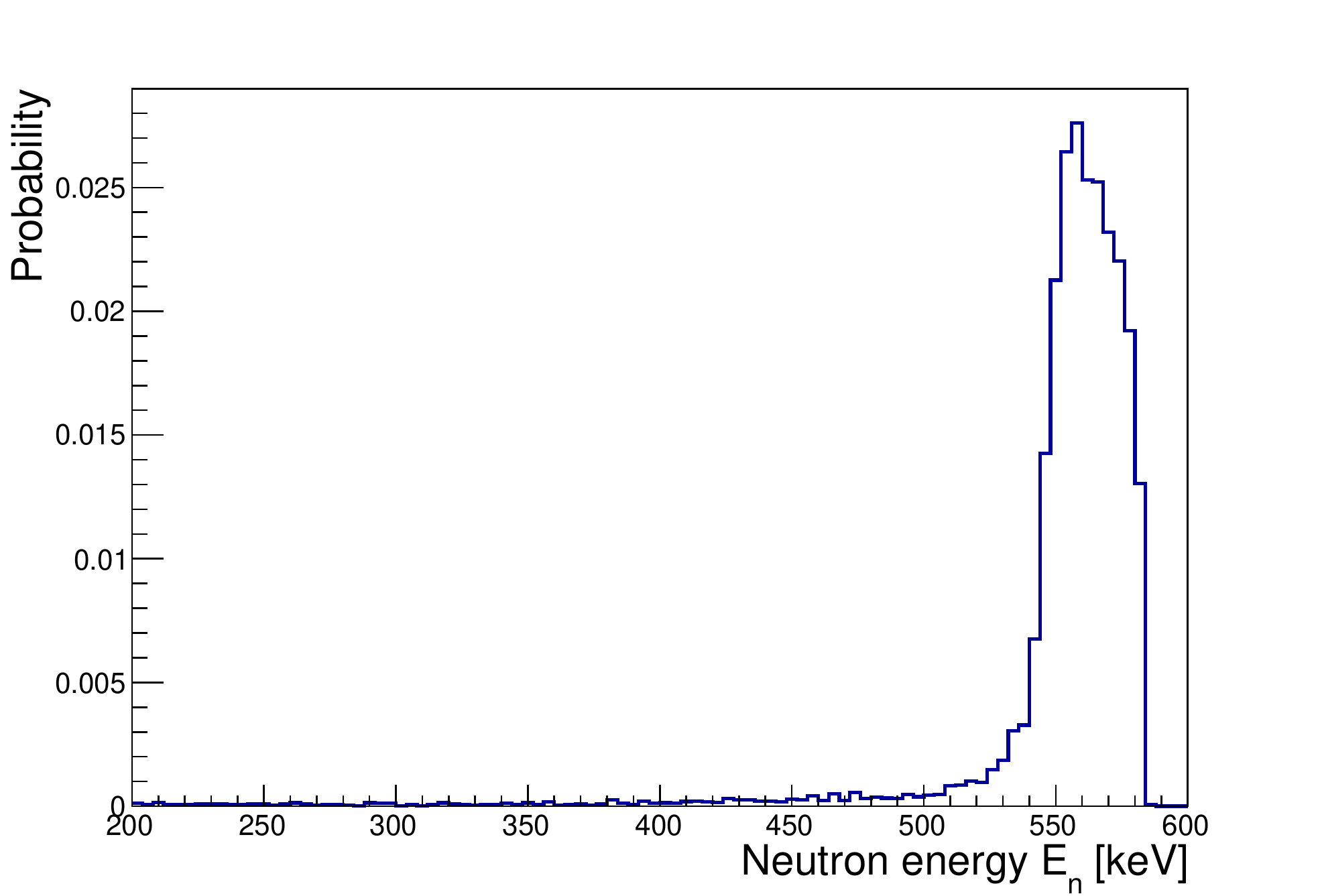}
\caption{ Left panel: simulation of the neutron spectrum interacting in the active volume of the detector placed at $D_{\mathrm{target}} = 30\ \mathrm{cm}$ and $\Theta = 0$. Neutron energies and direction distributions were estimated from the target using the TARGET software~\cite{Schlegel2005} and were propagated using MCNPX~\cite{Waters2002} }
		\label{fig:SimuSpectrumNeutron}
\end{figure}

In general, as neutrons are the ultimate background for DM detection, these particles can be used for mimic a WIMP signal in DM detectors. 
In order to evaluate the MIMAC neutron detection response and its low energy electron/recoil discrimination,
a mono-chamber MIMAC detector with an 18~cm drift space was placed in a monochromatic neutron field.
It was generated by the AMANDE (Accelerator for Metrology And Neutron Applications for External Dosimetry) facility~\cite{Gressier2003}
at the IRSN of Cadarache by using a $(p,n)$ nuclear reaction on a thin $^{7}\mathrm{Li}$ target ($140\ \mathrm{\mu g/cm^2}$) on an $\mathrm{Al\,F_3}$ backing.

This $(p,n)$ nuclear reaction has many resonances one of them generating a neutron field with a maximum kinetic neutron energy of $565\,\mathrm{keV}$ at $\Theta =0$. Here, $\Theta$ denotes the angle between neutron direction and the proton beam: kinetic energy of neutrons depends on $\Theta$ angle. 

The MIMAC mono-chamber ($10\times10\time18\mathrm{cm^3}$) detector were placed at $\Theta =0$ and at 30~cm distance. In this configuration, the solid angle covered by the detector is $\Omega = 0.111\ \mathrm{sr}$, thus neutron energy variations are small in the active volume and the neutron field can be considered as monochromatic, \textit{i.e.} mono-energetic. In order to confirm this hypothesis, a Monte-Carlo model of the neutron production and propagation was developed. Angular and energy distribution of neutrons outgoing from the target were estimated using TARGET software~\cite{Schlegel2005} and neutrons were propagated by Monte Carlo using MCNPX~\cite{Waters2002} considering the full geometry of the detector and the experimental hall. 
Figure~\ref{fig:SimuSpectrumNeutron} presents the Monte-Carlo simulation of the kinetic energy spectrum of neutrons interacting in the active volume showing
the mean energy of the monochromatic neutron field produced at 565~keV with an energy resolution of $\Delta E_n/E_n = 3\%$. The tail of the distribution corresponds to the backscaterred neutrons. In conclusion, the neutron field can be considered as a monochromatic neutron field.

In addition, neutron production is going along with an important $\gamma$-rays background from $(p,\gamma)$ reaction on $\mathrm{^7Li}$ and $\mathrm{^{19}F}$. Indeed, proton bombardment of $\mathrm{^7Li}$ and $\mathrm{^{19}F}$ produces high energy $\gamma$-rays lines (from $\sim15$ to $\sim18\ \mathrm{MeV}$ for $\mathrm{^7Li}$ and from $\sim6$ to $\sim7\ \mathrm{MeV}$ for $\mathrm{^{19}F}$ ) in $4\pi$~\cite{PhysRev74315}. The relative amplitude of these lines depends on the proton beam energy. 

\begin{figure}[tbp] 
\centering

		\includegraphics[width=0.8\linewidth]{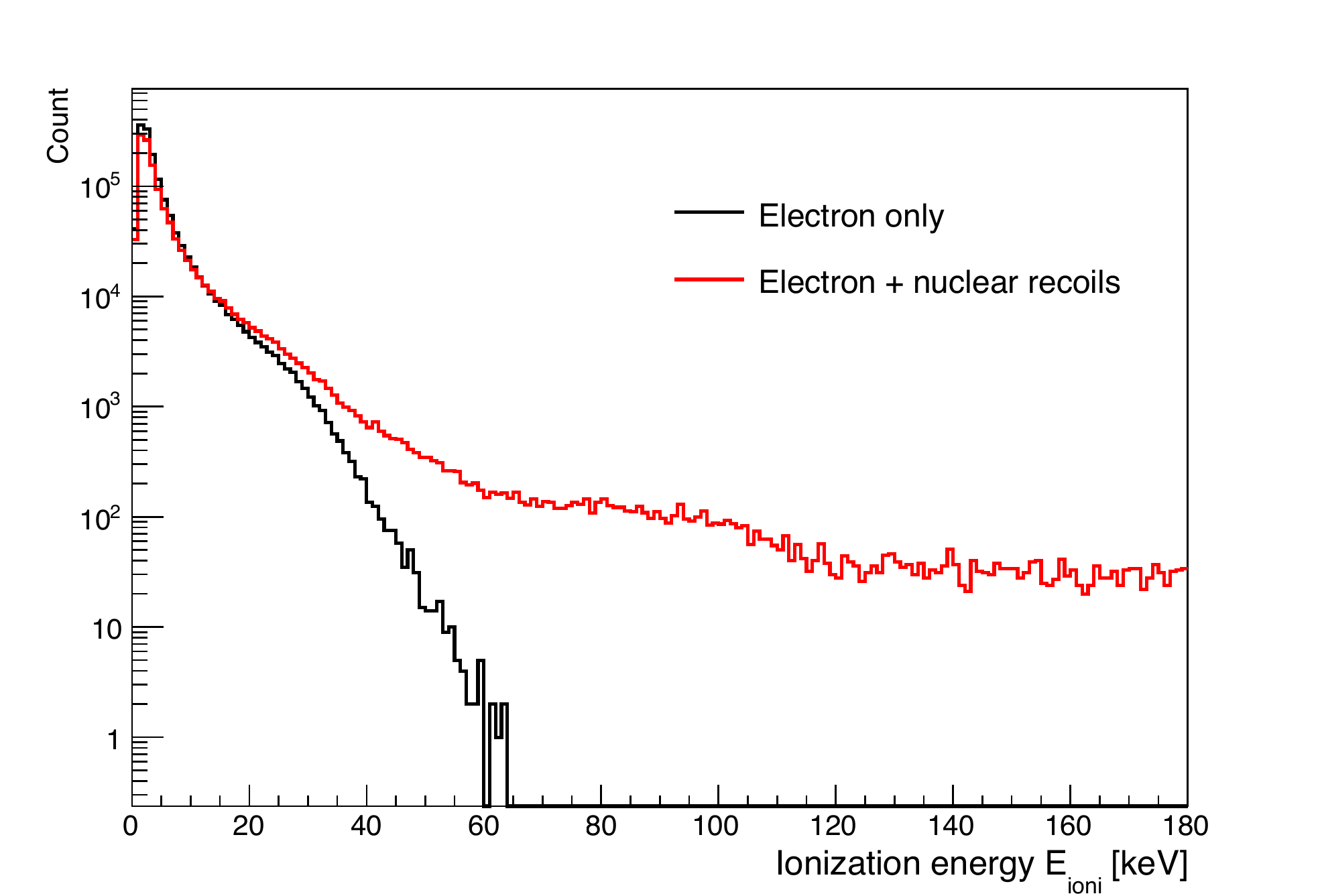}
		\caption{ Raw MIMAC energy spectra measured at the AMANDE facility. The black line corresponds to the target without \Nucl{Li}{7} (only electron events) and the red one to the target with the \Nucl{Li}{7} (electron and nuclear recoil events).}
		\label{fig:RawData}
\end{figure}

\subsubsection{Fast neutron detection}

Neutron elastic scatterings on nuclei in the active volume produce nuclear recoils with kinetic energies ranging from 0 to the maximum transferred kinetic energy depending on the nucleus mass, the so-called end-point. 565~keV neutrons transfer up to 107~keV in kinetic energy to \Nucl{F}{19} recoils.
However, for nuclear recoils there is a difference between the measurable ionization energy and the kinetic energy which is parametrized by the ionization quenching factor (IQF). This difference increase as the kinetic energy decreases. The IQF depends, in addition, on the nuclear recoil mass and gas properties (composition, pressure, temperature and impurities) and it can be estimated using the SRIM simulation code or measured as proposed in~\cite{Guillaudin2012}. 
In the case of fluorine, taking into account an estimation of the IQF from our measurements~\cite{Guillaudin2012}, a 107~keV nuclear recoil should release in ionization roughly 57~keV.  

The neutron production method using the $^7$Li(p,n) nuclear reaction produces an important $\gamma$-ray background from $(p,\gamma)$ channels on the Li target and on the fluorine of the $\mathrm{AlF_3}$ backing.
These $\gamma$-rays induced a huge number of electron recoils mainly by Compton scattering in the detector vessel, field cage and gas volume.

In order to evaluate the electron event rejection, data-taking with and without \Nucl{Li}{7} on the target were performed. In the first case, we have a neutron production along with an important $\gamma$-ray production from the $\mathrm{AlF_3}$ backing. In the second case, with only the $\mathrm{AlF_3}$ backing, only $\gamma$-rays are produced.
Figure~\ref{fig:RawData} shows the raw energy spectra measured by the MIMAC chamber at the AMANDE facility with (red line) and without (black line) \Nucl{Li}{7} on the target.
Both spectra present quite the same shape below 30~keVee. On the "$\gamma\mbox{-rays}$ only" spectrum (black line), we can see that the ionization energy released by electrons in the active volume does not exceed 60~keVee. This is due to the combined effects of detector geometry, the low electronic stopping power density and the long tracks of high energy electrons at 50 mbar. The raw "$n+\gamma\mbox{-rays}$ spectrum" shape makes even difficult to identify the two end-points from fluorine and carbon. The proton end-point at 565~keV is out of the Flash-ADC range.

\section{Discriminating observables}

\begin{figure}[tbp]
\centering 
		\includegraphics[width=0.49\linewidth]{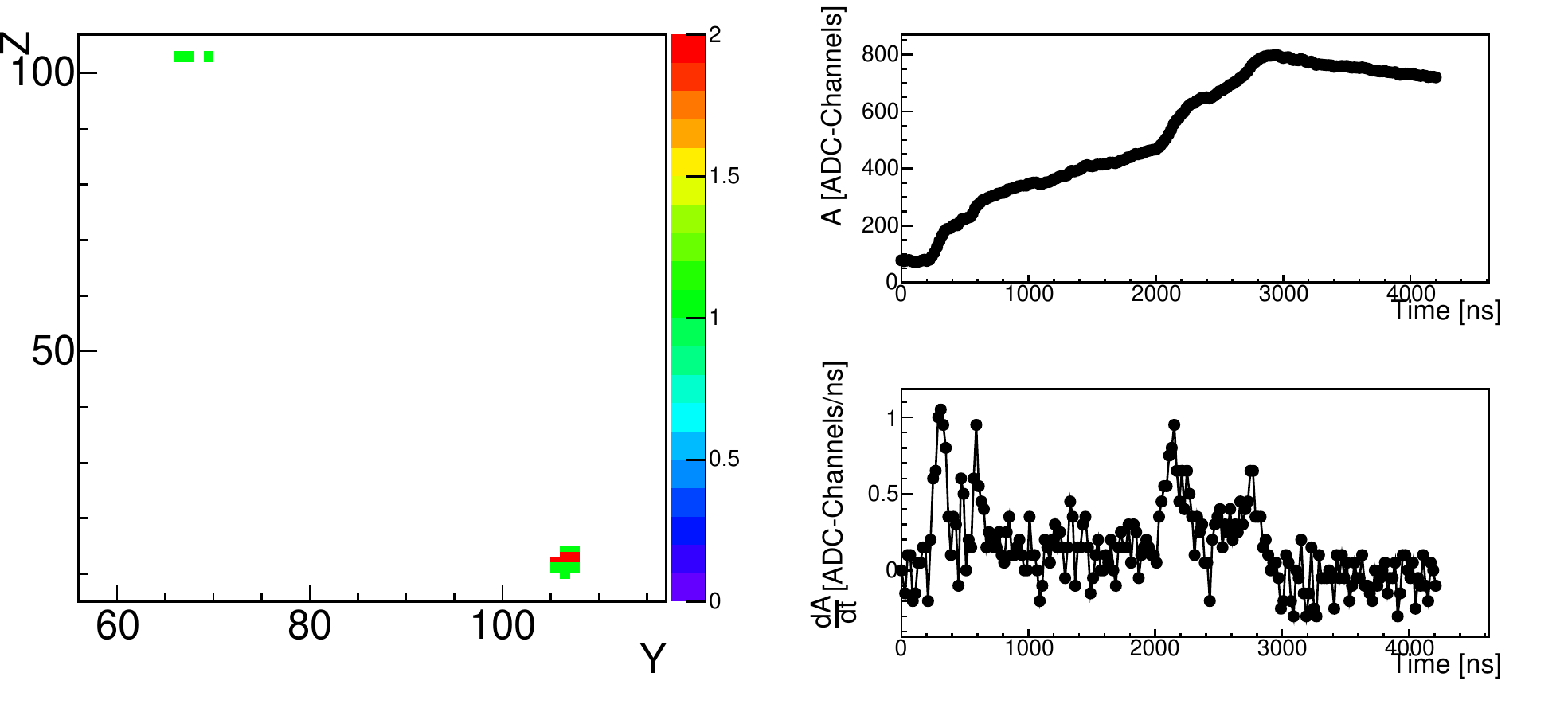}		
		\includegraphics[width=0.49\linewidth]{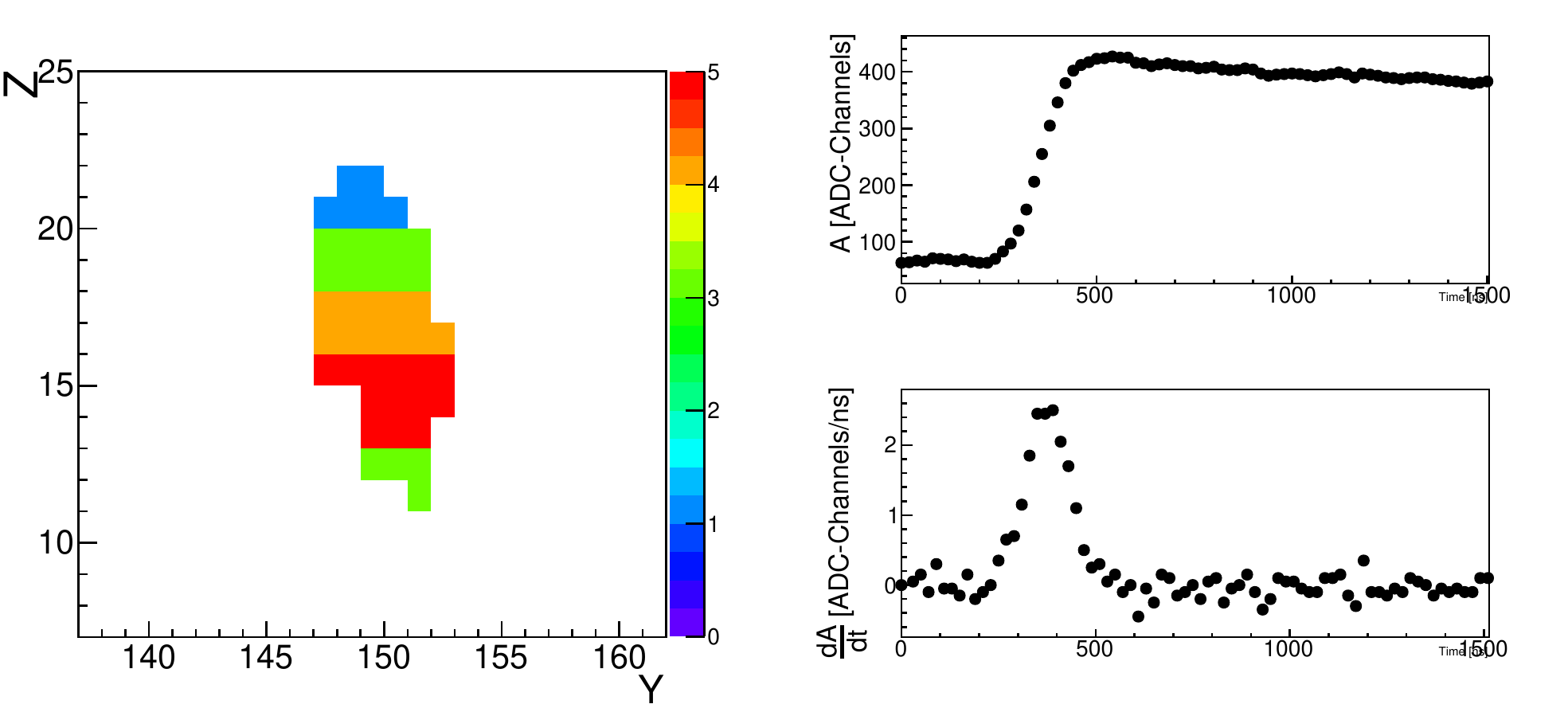}
		\caption{ Left and right panels present respectively the $(Y,Z)$ projection, the charge integrator amplitude and its first derivative of a 36.8~keV electron track and a 20.5~keV nuclear recoil track. The Z axis is in units of time-slices (20 ns each) and the Y axis in strip numbers. The colour scale corresponds to the relative number of strips fired in each time-slice.}
		\label{fig:Track}
\end{figure}

The charge integrator amplitude is continuously read by the MIMAC electronic read-out. An event is acquired recording both strips of pixels and grid information only if the following condition is fulfilled:
\begin{equation}
	A[i] - A[i-16] > E_{th}\,\, (\mbox{ADC units})
	\label{eq:trig}
\end{equation}
where $A[i]$ is the preamplifier amplitude in the $\mathrm{i^{st}}$ 20~ns time-slice and $E_{th}$ the threshold value. 
Figure~\ref{fig:Track} presents typical electron (left panel) and nuclear recoil (right panel) track projections on the $(Y,Z)$ plane associated to their preamplifier signals and its first derivative. This figure highlights the differences on the track topology and on their pulse-shape.
The electron event $(Y,Z)$ projection shows two small pixel clusters instead of the nuclear recoil event projection showing only one big cluster with a well-defined spatial development. The electronic event profile presents several "jumps" associated to charge clusters, while the nuclear recoil event presents only one "jump" with a faster rise-time.

For directional DM search, 3D tracks reconstruction of nuclear recoils~\cite{Billard2012c} has to be performed  to extract the track direction in the galactic rest frame. In order to get this 3D track determination, nuclear recoils have to be discriminated from electron and gamma background taking advantage of the electron and nuclear recoil event differences illustrated in Figure~\ref{fig:Track}. 

As a first step, to reject the active volume out(in)-going and miss reconstructed events some minimal cuts have to be applied.
In a second step, an electron/recoil discrimination based on track topology and signal pulse-shape observables will be applied.

\subsection{Minimal cuts}
\label{sec:Minimal}

 The minimal cuts applied to reject the mis-reconstructed events and out(in)-going events are the following:

	\textbf{Track}. Primary electron ionization densities of electron tracks are often not sufficiently high to trigger the strips of pixels in one 20 ns time-slice. A first cut consists of requiring events with a 3D track \textit{i.e.} more than one strip of pixels in coincidence (X and Y).

	\textbf{Out(in)-going events}. We define an active volume on the (X,Y) projection to the anode in order to reject all out(in)-going events. For these events, only a part of their track is included in the active volume and the energy measurement will be misestimated. 
	
	\textbf{Clustering}. The ionization (electron-ion) pair distribution produced by a nuclear recoil is denser, per length unit, than the distribution produced by an electron of the same energy even taking into account the {\it IQF}. This is due to the fact that the total integrated stopping power of a nuclear recoil is higher than electron one at the same energy mainly from its much shorter track. 
 We define a track cluster by a set of contiguous strips of pixels fired during a number of 20 ns time-slices. A nuclear recoil event will present only one cluster instead of the electron events presenting in general more than one.  
We will reject events presenting more than one cluster. Only those with two clusters separated in the X-Y projection by only one strip of pixels will be accepted.

\begin{table}[tbp]
\centering
\begin{center}
\begin{tabular}{lcc}
\hline
Cuts & Without $\mathrm{Li}$ & With $\mathrm{Li}$ \\
\hline
\hline
None & 893779 & 795596 \\
\hline
 Track & 78910 & 154603 \\ 
Track + $(X,Y)$ fiducialisation & 77629 & 143855  \\
 Track + Cluster & 43605 & 105978  \\
 \hline
Track + $(X,Y)$ fiducialisation& \multirow{2}{*}{42979} & \multirow{2}{*}{99334}\\
 + Cluster& &  \\
 \hline
\end{tabular}
\end{center}
\caption{Detail of the impact of minimal cuts combination on the number of events in the two sets of data: without and with $\mathrm{Li}$ target.}
\label{tab:MinCutImpact}
\end{table}
\begin{figure}[tbp]
\centering 
		\includegraphics[width=0.7\linewidth]{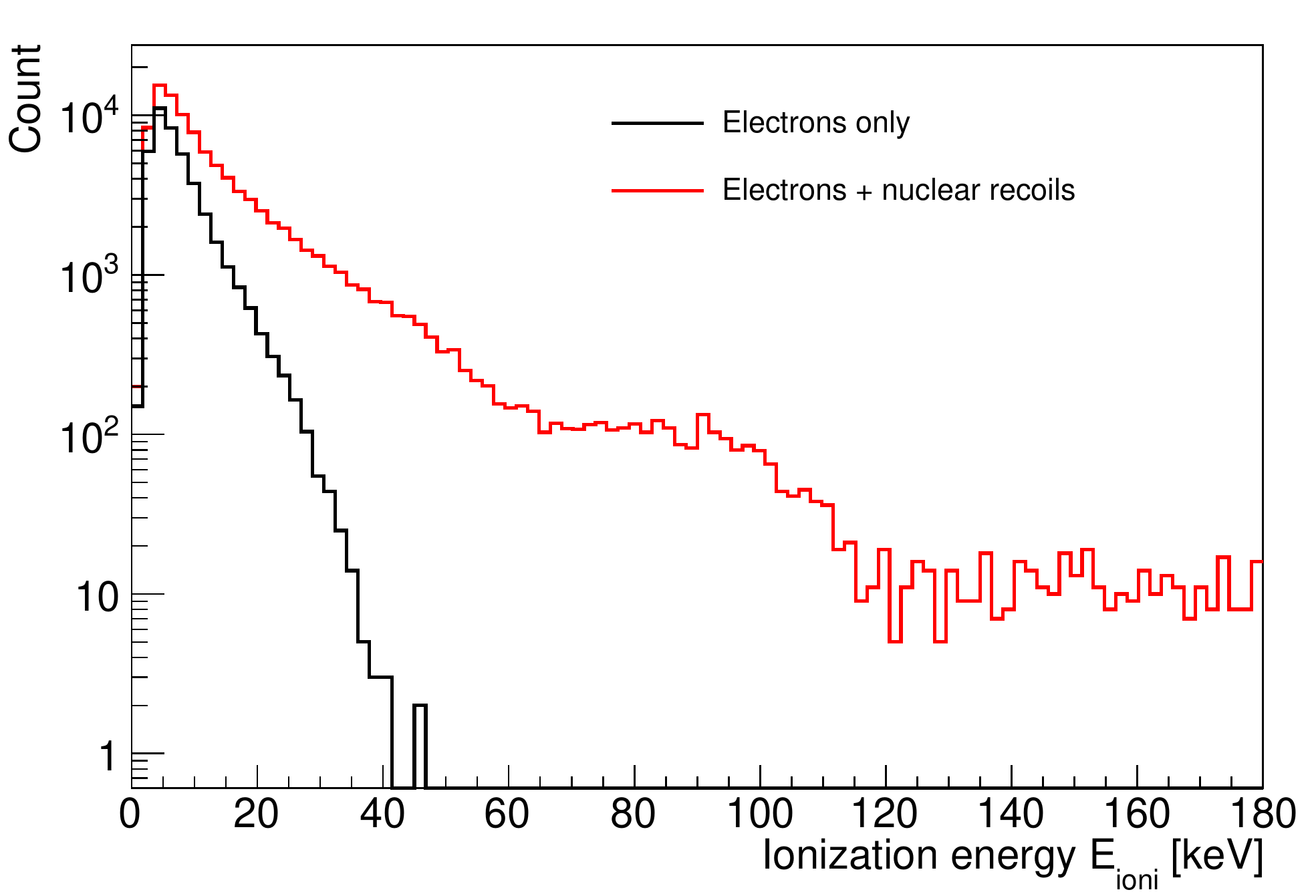}
		\caption{ Total energy spectra obtained after the application of minimal cuts. The black line corresponds to events detected with the target without \Nucl{Li}{7} (only $\gamma$-rays produced) and the red one to those detected with \Nucl{Li}{7} on the target ($\gamma$-rays and neutrons produced). }
		\label{fig:SpectreEMinimalCadarache}
\end{figure}

Table~\ref{tab:MinCutImpact} shows the impact of each minimal cut on the data sample used in the analysis.
Track requirement is the dominant cut and we can note that this cut has a higher effect on the electron only sample ($91\%$ reduction) than on the nuclear recoil and electron sample($80\%$ reduction). This difference comes from that the probability to fire the strip of pixels is lower for an electron than for a nuclear recoil.
In order to apply the two remaining cuts, tracks are required.
The application of the ($X,Y$) fiducialisation shows respectively a 1\% and a 7\% reduction of both samples. This difference comes from the fact that due to their ionization density, high energy electrons could not fire edge strips and cannot be identified as out/in-going event which is not the case for nuclear recoils.
The 45\% and 31\% reduction coming from the application of the cluster cut has the same origin.

Figure~\ref{fig:SpectreEMinimalCadarache} shows the energy spectra measured by the MIMAC chamber at the AMANDE facility with (red line) and without (black line) \Nucl{Li}{7} on the target. These spectra are obtained after the application of the minimal cuts described above.
On the "$n+\gamma\mbox{-rays}$" spectrum, \Nucl{F}{19} and \Nucl{C}{12} end-points at 57 and 110~keVee can clearly be identified.
These end-points define the maximum kinetic energies affected by the {\it IQF}. In the case of \Nucl{F}{19} the {\it IQF} was measured at 46\% at 50 keV kinetic energy by the method proposed in~\cite{Guillaudin2012}, showing that the end point at 57~keV in ionization is consistent with our measurements.
 The \Nucl{H}{1} end-point at 565~keV even affected by the quenching is out of range.

\subsection{Discriminating observables}
\label{sec:ObsBDT}

\begin{figure}[tbp]
\centering 
		\includegraphics[width=0.8\linewidth]{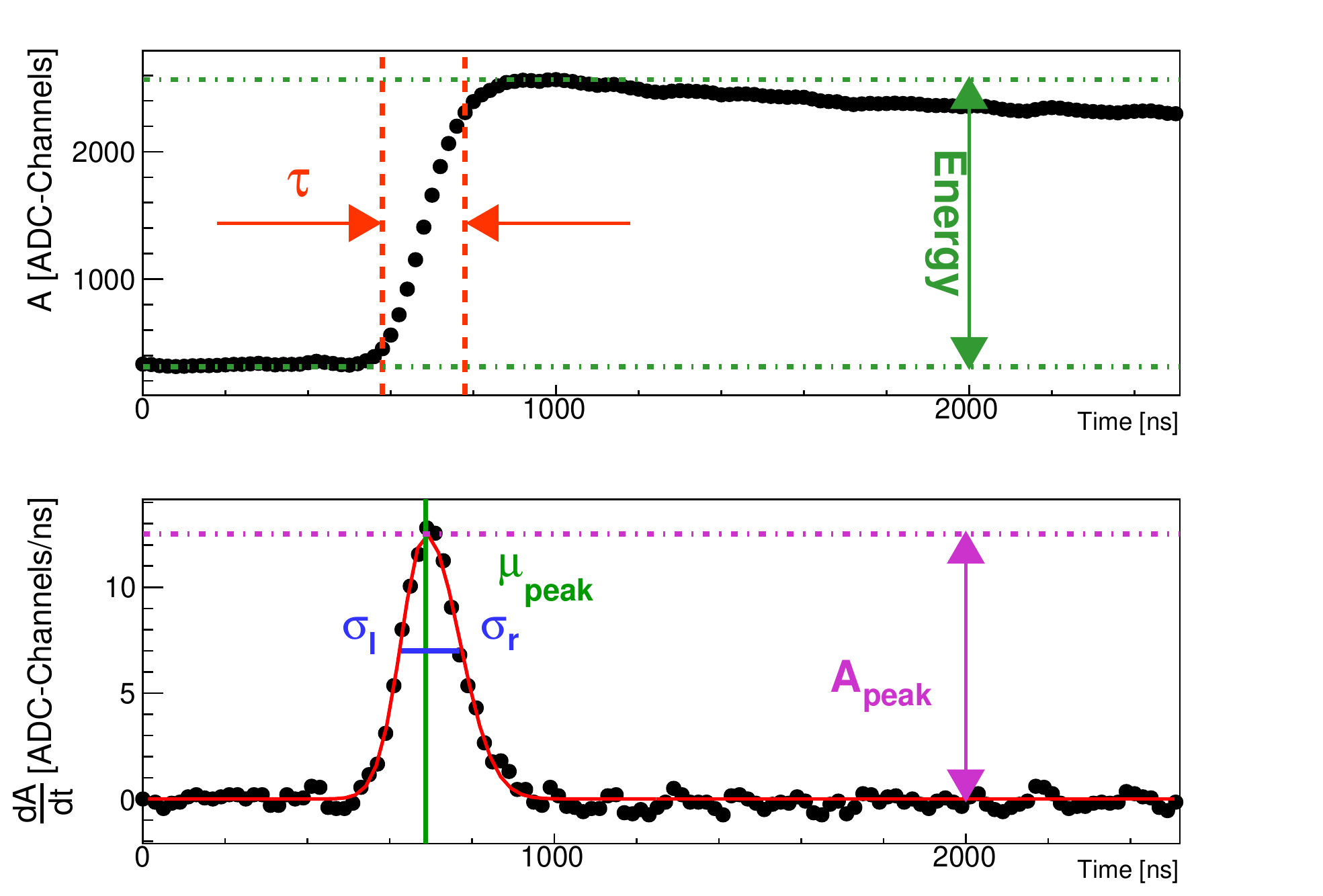}
		\caption{Top panel: A 38.3~keVee event preamplifier amplitude as a function of time. 
			     The green arrow represents the ionization energy. 
			     The red arrows and dashed lines represent the rise-time definition. Bottom panel: The preamplifier amplitude first derivative as a function of time. The red line shows the fit with an asymmetric gaussian function. The purple, green and blue lines presents the fit parameters.}
		\label{fig:Flash}
\end{figure}

\begin{figure*}[tbp]
\centering 
	\includegraphics[height=0.85\linewidth,angle=90,origin=c]{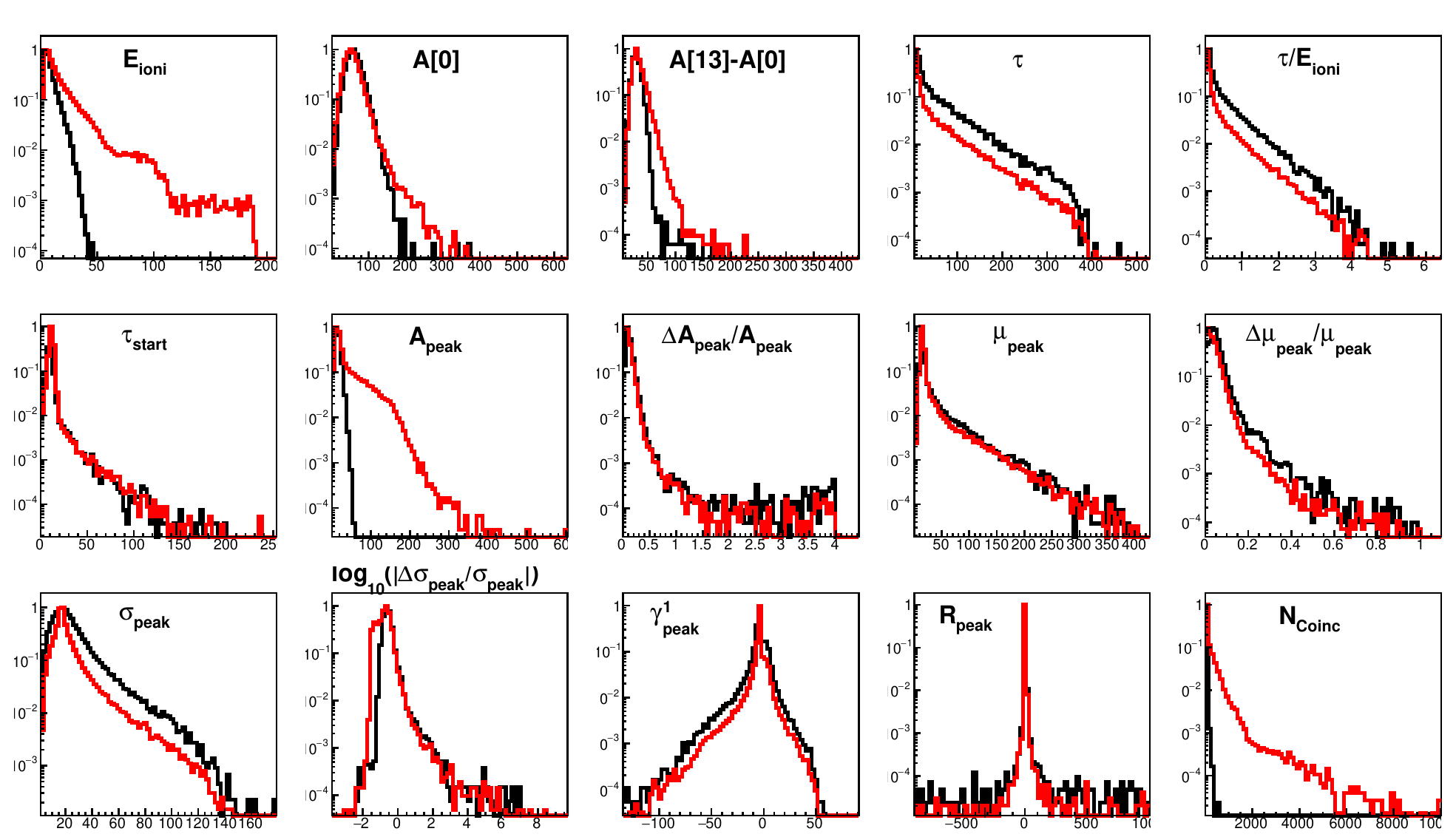}
	\caption{ One-dimension distribution of each discriminating observables (part 1/2). The black line corresponds to the events detected with the target without \Nucl{Li}{7} (only $\gamma$-rays) and the red one to those detected with \Nucl{Li}{7} ($\gamma$-rays and neutrons). }
	\label{fig:DiscriVar_1}
\end{figure*}

\begin{figure*}[tbp]
\centering 
	\includegraphics[height=0.85\linewidth,angle=90,origin=c]{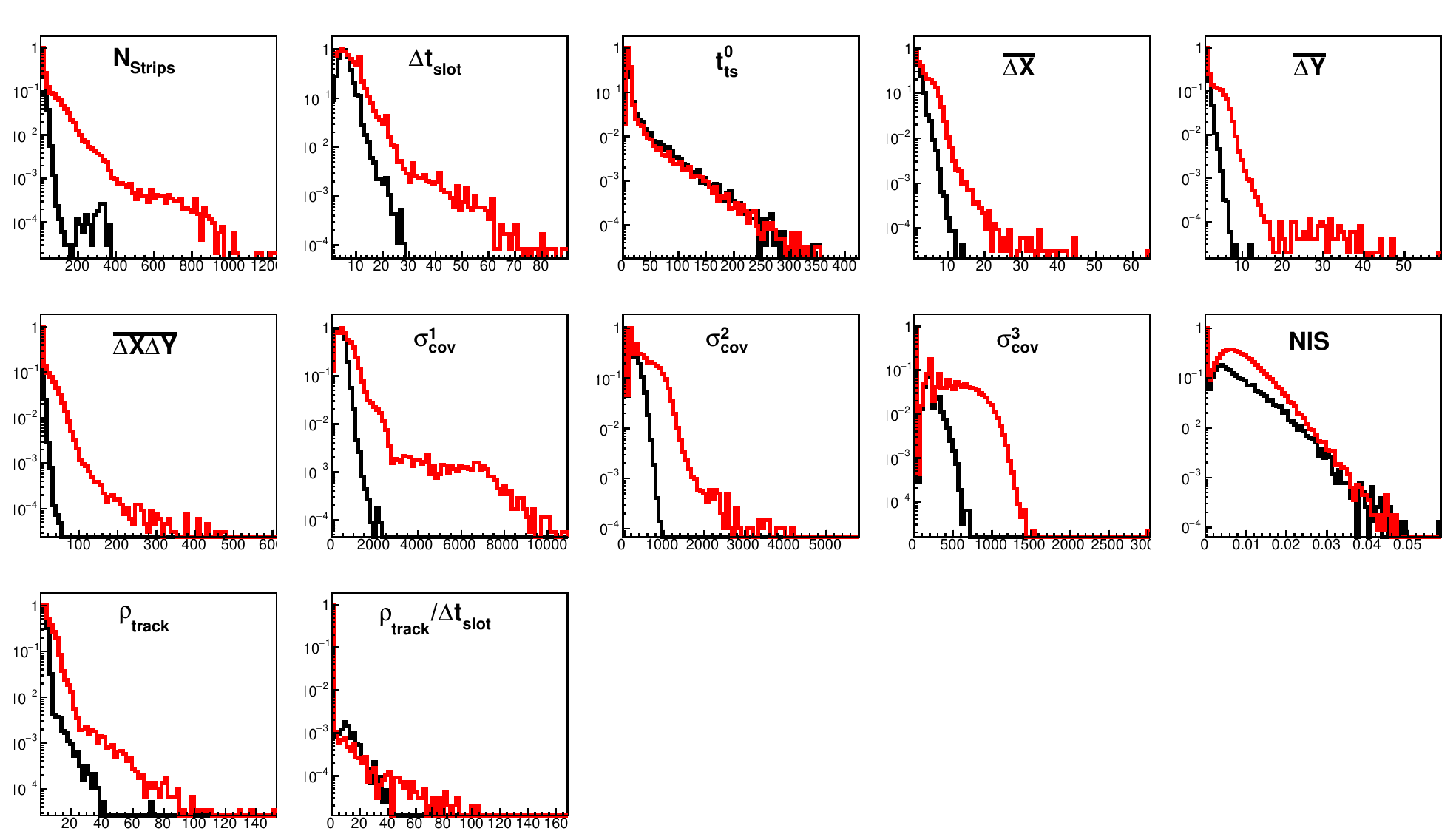}
	\caption{One-dimension distribution of each discriminating observables (part 2/2). The black line corresponds to the events detected with the target without \Nucl{Li}{7} (only $\gamma$-rays) and the red one to those detected with \Nucl{Li}{7} ($\gamma$-rays and neutrons). }
	\label{fig:DiscriVar_2}
\end{figure*}

Using both, the charge preamplifier profile signal and the track topology, several observables are defined to discriminate nuclear recoil events from electron recoil ones. We can distinguish two kinds of observables: pulse-shape and track topology observables. Figures~\ref{fig:DiscriVar_1}~and~\ref{fig:DiscriVar_2} present the one-dimension distribution for each observable. Black line corresponds to the events detected with the target without \Nucl{Li}{7} (only $\gamma$-rays) and the red one to those detected with \Nucl{Li}{7} ($\gamma$-rays and neutrons)

The pulse-shape is directly related to the primary electron-ion pairs distribution shape in the active volume. Using the fast pre-amplifier response (roughly 60~ns rise-time) several observables are defined. Figure~\ref{fig:Flash} shows a 38.3~keVee preamplifier amplitude as a function of time and its first derivative illustrating some observable definitions. 

\textbf{Ionization Energy ($\mathbf{E_{ioni}}$}). The ionization energy is defined as the difference between the maximum and minimum preamplifier signal amplitudes. Top panel of figure~\ref{fig:Flash}, illustrates the energy measurement from a typical flash-ADC signal.

\textbf{Offset ($\mathbf{A[0]}$)}. As the anode is continuously read by the electronics, before triggering, an event could have a residual charge coming on the grid from an event that has not had enough charge to trigger the preamplifier. This residual charge coming before the event in the preamplifier is called the offset. In the case of an electron recoil, the ionization density along the z-axis could be not enough to trigger the charge preamplifier.

\textbf{Preamplifier first derivative fit parameters ($\mathbf{A_{peak}}$, $\mathbf{\Delta A_{peak}/A_{peak}}$, $\mathbf{\mu_{peak}}$, $\mathbf{\Delta\mu_{peak}/\mu_{peak}}$, $\mathbf{\sigma_{peak}}$, $\mathbf{\log_{10}(\Delta\sigma_{peak}/\sigma_{peak})}$ , $\mathbf{R_{peak}}$, $\mathbf{\gamma^1_{peak}}$ )}.
	The peak in the first derivative of the preamplifier signal is fitted using an asymmetric gaussian function ({\it i.e. $\sigma_l (x<\mu) \neq \sigma_r (x>\mu) $}). 
	From this fit four parameters are extracted: the amplitude $A_{peak}$, the time position $\mu_{peak}$, the left half-width $\sigma_l$ and the right half-width $\sigma_r$.
	From the last two parameters, we can define an asymmetry factor $R_{peak} = \sigma_l/\sigma_r$ associated to the charge collection. In addition the reduced ${\chi^2}_{\mathrm{peak}}$ is calculated.

\begin{figure}[tbp]
\centering 
		\includegraphics[width=0.7\linewidth]{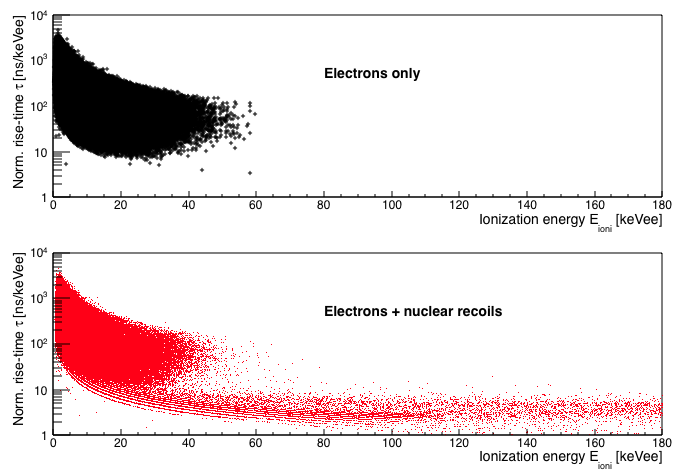}
		\caption{Event distributions on the plane $(E_{\mathrm{ioni}},\tau/E_{\mathrm{ioni}})$. 
			     The black dots correspond to the events detected with the target without \Nucl{Li}{7} (only $\gamma$-rays produced) 
			     and the red one to those detected with \Nucl{Li}{7} on the target ($\gamma$-rays and neutrons produced) }
		\label{fig:RiseTime}
\end{figure}

\textbf{Rise-Time and normalized Rise-Time ($\mathbf{\tau}$, $\mathbf{\tau_{start}}$ and $\mathbf{\tau/E_{ioni}}$ }).	
The rise-time is defined as the time elapsed between 10~\% and 90~\% of the maximum amplitude of the preamplifier signal. This rise-time depends, obviously, on the event ionization energy. In order to define a discriminating observable, we normalize it by the total ionization energy ${\tau/E_{ioni}}$. 
Figure~\ref{fig:RiseTime} presents the event distributions in the $(E_{\mathrm{ioni}},\tau/E_{\mathrm{ioni}})$ plane for $n+\gamma\mbox{-rays}$ (red dots) and $\gamma\mbox{-rays}$ only (black crosses). This figure shows that the nuclear recoil normalized rise-time is systematically lower than the electron normalized rise-time. 
Moreover, we define the start rise-time $\tau_{start}$ as the time when the preamplifier amplitude is higher than 10~\% of the maximum amplitude.

\begin{figure}[tbp]
\centering 
	\includegraphics[width=0.8\linewidth]{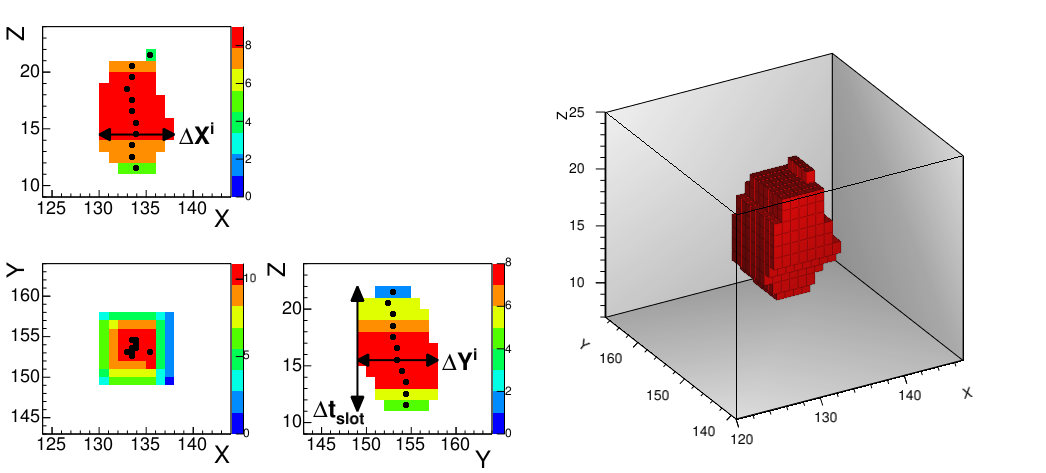}
	\caption{Right panels: projections of a 38.3~keVee nuclear recoil track in the $(X,Z)$, $(Y,Z)$ and $(X,Y)$ plans.
	              The $Z$ axis is in units of time-spice (20 ns) and the $X$ axis in strip number. 
	              The black dots represent the time-slice barycenter position.
	              The vertical arrow represents the definition of the track duration. 
	              The horizontal arrow represents one time-slice width $\Delta X$ along the X axis.
	              The colour scale corresponds to the relative number of strips fired in the time-slice.}
	\label{fig:Topo}
\end{figure}

The pixelated Micromegas coupled to the fast electronics provide a sampling of the $(X,Y)$ ionization electron density as a function of time. As previously mentioned, by knowing the electron drift velocity, a 3D track could be reconstructed. 
Figure~\ref{fig:Topo} shows projections of a 38.8~keVee track on $(X,Z)$, $(Y,Z)$, $(X,Y)$ planes and its 3D reconstruction. In this figure, the graphic representation of $\Delta X_i$, the width of the $i^{st}$ time-slice along the $x$ axis, is also shown.
Then, using this information, we define a set of track observables which some of them are illustrated in figure~\ref{fig:Topo}.

\textbf{Track duration and track start ($\mathbf{\Delta t_{slot}}$ and $\mathbf{t^0_{track}}$)}. Figure~\ref{fig:Topo} illustrates the definition of the track duration $\Delta t_{slot}$. 
The track duration is the difference between the last time-slice $(t_{track}^{end})$ and the first one $(t_{track}^{0})$ as shown in figure~\ref{fig:Topo}. This observable is related to the projection of the track length along the $z$-axis. 
On the other hand, $t^0_{track}$ is the time-slice number of the first strip coincidence. The shift between the trigger and $t^0_{track}$ is related to the ionization electron density. In the case of an electron recoil, the value of $t^0_{track}$ may fluctuate due to the low ionization density. In contrast, in the case of nuclear recoils this shift is expected to be more or less constant.

\textbf{Strip and coincidence number $\mathbf{N_{strips}}$ and $\mathbf{N_{coinc}}$ }. 
The strip number and coincidence number correspond respectively to the total number of strips and (X,Y) coincidences triggered during the event. If the full primary electron ionization density is detected, these two quantities are expected to be linearly correlated. Nuclear recoil ionization density is sufficiently important to trigger strips and (X,Y) coincidence in one 20 ns time-slice, whereas for electron recoils this is not the case.

	\textbf{Normalized Integrated Straggling (NIS)}.
	The NIS is defined as the sum of each barycenter deviation $\Delta\theta_i$ along the track and normalized by the ionization energy: 
	\begin{equation}
		NIS = \frac{1}{E_{ioni}}\sum_{i=1}^{N_s-2}\Delta\theta_i
	\end{equation}
	This observable estimates the integrated straggling along the track. The straggling depends on the recoil mass and gas pressure. The NIS of the electrons will be larger than the NIS of nuclear recoils of the same kinetic energy~\cite{Billard2012b}.

\begin{figure}[tbp]
\centering 
		\includegraphics[width=0.7\linewidth]{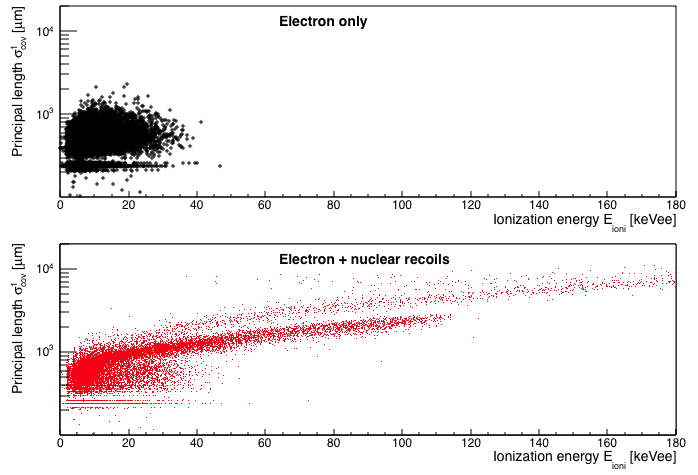}
		\caption{ Event distributions on the plane $(E_{\mathrm{ioni}},\sigma_{cov}^1)$.
			      The black dots correspond to the events detected with the target without \Nucl{Li}{7} (only $\gamma$-rays) 
			      and the red ones to those detected with \Nucl{Li}{7} ($\gamma$-rays and neutrons) on the target.
			      }
		\label{fig:DistribExTrack}
\end{figure}

\begin{figure}[tbp]
\centering 
		\includegraphics[width=0.99\linewidth]{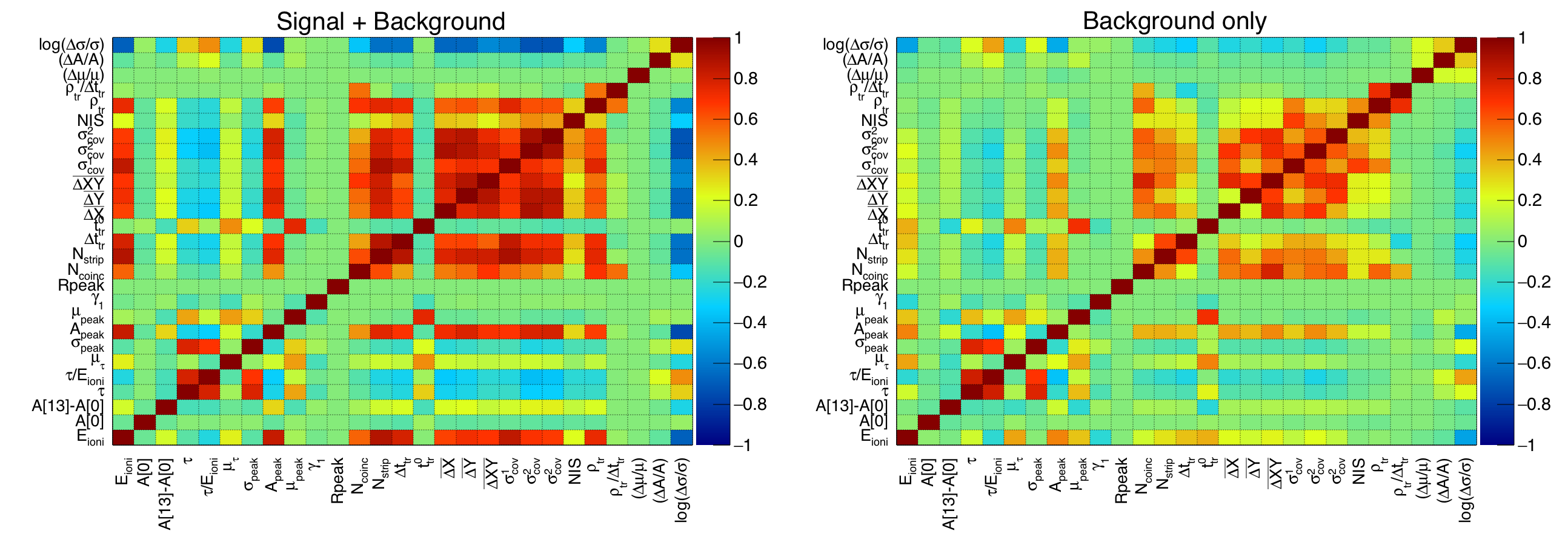}
		\caption{ The left and right panels present the correlation matrices of the MIMAC observables used for the Boosted Decision Tree (BDT) analysis. The left panel corresponds to the measurement with the \Nucl{Li}{7} ($\gamma$-rays and neutrons) and the right panel to the measurement without \Nucl{Li}{7} (only $\gamma$-rays).}
		\label{fig:Correlation}
\end{figure}

	\textbf{Track density and normalized track density ($\mathbf{\rho_{track}}$ and $\mathbf{\rho_{track}/\Delta t_{slot}}$ )}. This observable is related to the primary ionization electron density. It is defined as:
	\begin{equation}
		\rho_{track} = \sum_{i=1}^{N_{s}}{\frac{N_{pix}^{i}}{\Delta X^{i}\times \Delta Y^{i}}}
	\end{equation}
	where $N_{s}$ is the number of time-slices, $N_{pix}^{i}$ is the number of pixels fired in the $i^{\mathrm{st}}$ time-slice and $\Delta X(Y)^{i}$ the width on the $X(Y)$ axis in the $i^{\mathrm{st}}$ time-slice. The nuclear recoil track density will be, in general, higher than the electron track density due to the number of "holes" present in an electron track, see figure~\ref{fig:Topo}. Moreover, we defined the normalized track density as $\rho_{track}/\Delta t_{slot}$. 

\textbf{Track widths ($\overline{\Delta X}$, $\overline{\Delta Y}$ and $\overline{\Delta X\Delta Y}$)}. From the $\Delta (X/Y)^{i}$ time-slice width, we calculate the mean of $\Delta X$, $\Delta Y$ and $\Delta X\Delta Y$. 
Mean value of $\Delta X$ and $\Delta Y$ are related to the track length projected along the $X$ and $Y$ axis. In the cas of tracks almost contained in the $(X,Y)$ plane, these observables are sensitive to the track length:  in comparison with a nuclear recoil, lower value of $\overline{\Delta X}$ and $\overline{\Delta Y}$ are expected for an electron recoil. 
The $\overline{\Delta X\Delta Y}$ observable corresponds to the mean surface of the track on the anode plane, the same behavior is expected for this observable.

\textbf{Track principal component length and widths ($\sigma^1_{cov}$, $\sigma^2_{cov}$ and $\sigma^3_{cov}$)}.
As shown in the right panel of figure~\ref{fig:Topo} we can reconstruct a 3D track from the MIMAC read-out. From this information and using a principal component analysis, we can calculate the longitudinal track length $\sigma^1_{cov}$ and its transverse widths $\sigma^2_{cov}$ and $\sigma^3_{cov}$. 
These lengths are the eigenvalues of the track position covariance matrix.
The longitudinal track length observable $\sigma_{Long}$ is another estimator of the track length. 
These observables are related to the electron diffusion in the drift space and to the track direction. Indeed, for a nuclear recoil event an X/Y asymmetry is expected from the track direction.
However, this track "fitting" approach is not adapted for direction extraction for low energy recoils $(E_{ioni}< 40\,\mathrm{keV}$). A more complex method is needed in order to determine the track direction with the MIMAC read-out. A dedicated paper~\cite{Billard2012c} has proposed an original likelihood method based on track simulations for low energy tracks.

		Figure~\ref{fig:DistribExTrack} represents the event distributions in the $(E_{\mathrm{ioni}},\sigma_{cov}^1)$ plan for $n+\gamma\mbox{-rays}$ (red dots) and $\gamma\mbox{-rays}$ only (black crosses). Two different regions corresponding to electrons and nuclear recoils can be identified. Moreover, on the $n+\gamma\mbox{-rays}$ distribution two branches can be identified, corresponding the shorter one to fluorine and carbon recoils and the longer one to proton recoils.

\begin{figure}[tbp]
\centering 
		\includegraphics[width=0.7\linewidth]{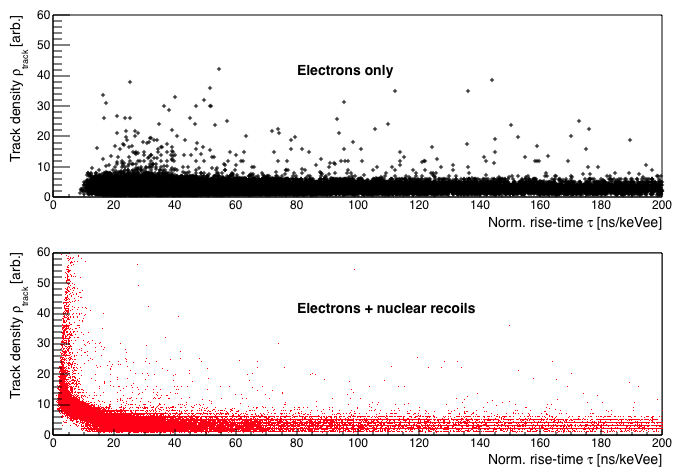}
\caption{ Event distributions on the plans $(\tau/E,\rho_{\mathrm{track}})$. The black dots correspond to the events detected with the target without \Nucl{Li}{7} (only $\gamma$-rays) and the red ones to those detected with \Nucl{Li}{7} ($\gamma$-rays and neutrons) on the target. 
			      }
		\label{fig:DensityDoublek}
\end{figure}

Figure~\ref{fig:Correlation} shows the correlation matrices of the observables defined above for the $n+\gamma$-rays (left panel) and $\gamma$-rays only (right panel) samples. These correlation matrices illustrate the previous observation. Indeed,  
in the $n+\gamma$-rays sample, the $\rho_{track}$ and $\tau/E$ observables are anti-correlated (-25\%) while in the $\gamma$-rays only sample, these observables are almost uncorrelated (-3\%). 
Figure~\ref{fig:DensityDoublek} shows the event distributions in the $(\tau/E,\rho_{track})$ plan for $n+\gamma\mbox{-rays}$ (red dots) and $\gamma\mbox{-rays}$ only (black crosses), illustrating the previous correlation values. We can clearly identify two different regions corresponding to electron and nuclear recoils. This figure illustrates the electron/recoil separation and the complementary of flash $(\tau/E)$ and track $(\rho_{track})$ observables. 
Moreover, $N_{Coinc}$ is correlated at 33 \% with the normalized track density ($\rho_{track}/\Delta t_{slot}$) for the $n+\gamma$-rays sample and it is anti-correlated at -27 \% for the $\gamma$-rays only sample.
These two examples illustrate how the observable combination will be used to differentiate both samples.

\section{Electron/recoil discrimination by boosted decision trees}

As demonstrated in~\cite{Billard2012b}, a sequential analysis of the electron/recoil discrimination is not sufficient to get a good discrimination power ($10^{4}-10^{5}$).  In this section, a boosted decision trees (BDT) analysis strategy and the results obtained will be presented.

\subsection{Boosted decision trees analysis strategy}

Boosted decision trees~\cite{Breiman1984} is a multivariate analysis algorithm widely used in high-energy physics. 
It can be seen as a data classifier, often employed for signal/background discrimination. 
It is based on the optimization of successive linear cuts on different discriminant observables.
The classification given by the BDT analysis is encoded on a BDT response variable defined as:
\begin{equation}
	X_{\mathrm{BDT}}  = \sum_{i=1}^{N_{\mathrm{trees}}}{\alpha_iT_i(\tilde{\mathcal{O}})}
\end{equation}
 where $N_{\mathrm{trees}}$ is the number of trees used for boosting, $\alpha_i$ the normalized weight of each tree $T_i$ and $\tilde{\mathcal{O}}$ the observables used in the analysis. By definition, the BDT variable value must be between -1 and 1.
 
In the particular case of the electron/recoil discrimination, the two following hypotheses are tested for each event:
\begin{equation*}
	\left\{
		\begin{array}{ll}
			H_0 =& \mbox{electron}\,(e^{-})\\
			H_1 =& \mbox{nuclear recoil}\, (R)
		\end{array}
	\right.
\end{equation*}
Using the data acquired with the fast neutrons produced at the AMANDE facility, it is not possible to obtain pure electron and/or nuclear recoil samples. The BDT analysis will be used to identify electron recoils on the nuclear recoil and electron sample by testing the two following hypothesis:
\begin{equation*}
	\left\{
		\begin{array}{ll}
			H'_0 =& \mbox{electron only ({\it i.e.} without \Nucl{Li}{7})} \\
			H'_1 =& \mbox{nuclear recoil + electron ({\it i.e.} with \Nucl{Li}{7})}
		\end{array}
	\right.
\end{equation*}
In conclusion, the BDT will be trained on AMANDE data set in order to separate electron recoils from the full data set acquired with $\mathrm{^{7}Li}$ target.

We applied a BDT analysis by using the TMVA software framework~\cite{Hoecker2007}.
We trained a forest of 2000 trees with $3.8\times 10^4$ events. In order to avoid the over-training while maximizing BDT performances,
we requested for each foil a minimum of $10\% \equiv 3.8\times 10^3 $ events and a maximal tree-level of 3. 
In order to evaluate the overtraining, we compared the train and the test samples using the Kolmogorov-Smirnov statistic test.
It tests if the train and the test samples follow the same probability distribution evaluating the maximal distance between the sample cumulative distributions.
We measure a $D = 3.11\times10^{3}$ maximal distance corresponding to a 0.996 p-value.
The confidence interval at $1\sigma$ is $[0;4.9\times 10^{-3}]$, it includes the maximal distance value. 
We can conclude that both samples follow the same probability distribution, validating our statement that our BDT analysis is not overtrained.

\begin{table}[tbp]
\centering
	\begin{tabular}{clcc}
		\hline
		\rule[-0.2cm]{0cm}{0.55cm} Rank & Variable & Importance & Type\\
		\hline
		\hline
		1 & $A_{peak}$                  			& 8.718e-02 	& Pulse-shape \\
		2 & $N_{Coinc}$                                		& 7.662e-02 	& Track 	\\
		3 & $\rho_{track}/\Delta t_{slot}$       	& 7.377e-02 	& Track 	 \\
		4 & $\Delta t_{slot}$                       		& 7.157e-02     	& Track 	 \\
 		5 & $N_{Strips}$                            		& 7.026e-02 	& Track 	\\
 		6 & $t_{tr}^{0}$                           		& 6.207e-02 	& Track \\
 		7 & $E_{ioni}$                            		& 4.464e-02 	& Pulse-shape \\
 		8 & $\tau$                            			& 4.395e-02 	& Pulse-shape \\
 		9 & $\mu_{peak}$                       		& 4.352e-02 	& Pulse-shape \\
 		10 & $A[0]$                              			& 3.620e-02 	& Pulse-shape \\
 		11 & $\sigma^3_{cov}$                      		& 3.593e-02 	& Track 	 \\
 		12 & $\tau/E_{ioni}$                     		& 3.563e-02 	& Pulse-shape \\
		13 & $A[13]- A[0]$                             	& 3.345e-02 	& Pulse-shape \\
		14 & $\sigma_{peak}$                        	& 3.254e-02 	& Pulse-shape \\
		15 & $\sigma^2_{cov}$                    		& 2.955e-02 	& Track 	  \\
		16 & $\overline{\Delta Y}$                  	& 2.909e-02 	& Track 	 \\
		 17 & $\sigma^1_{cov}$                    		& 2.779e-02 	& Track 	 \\
		 18 & $\rho_{track}$                           		& 2.671e-02 	& Track 	 \\
		 19 & $\overline{\Delta X}$                 	& 2.542e-02 	& Track 	\\
		 20 & $\log_{10}(\Delta\sigma_{peak}/\sigma_{peak})$    & 2.377e-02      & Pulse-shape  \\
		 21 & $\tau_{start}$                 			& 2.302e-02 	& Pulse-shape \\
		 22 & $\Delta A_{peak}/A_{peak}$     		& 1.948e-02 	& Pulse-shape \\	
		 23 & $NIS$                                 		& 1.932e-02 	& Track 	\\
		 24 & $\gamma^{1}_{peak}$                  	& 1.683e-02	& Pulse-shape  \\
		 25 & $\overline{\Delta X\Delta Y}$    	& 1.086e-02 	& Track 	\\
		 26 & $R_{peak} $                    			& 8.228e-04 	& Pulse-shape \\
		 27 & $\Delta\mu_{peak}/\mu_{peak}$       & 0.000e+00 	& Pulse-shape  \rule[-0.15cm]{0cm}{0.5cm} \\
		\hline
	\end{tabular}
	\caption{ BDT ranking of the twenty-seven discriminant observables of the Boosted Decision Tree analysis with their associated importance (see text for definition).}
	\label{tab:BDTImpotance}
\end{table}

The table~\ref{tab:BDTImpotance} presents the BDT ranking of the seventeen discriminant observables.
The ranking was established by calculating the importance of each variable~\cite{Hoecker2007}. 
The importance is evaluated as the total separation-gain weighted by the number of events for each variable.
It quantifies the importance of an observable in a BDT analysis. 
This table shows that there is no dominant observable involved in the separation and it illustrates the complementarity of the pulse-shape and the 3D track observables for electron/recoil discrimination.

\subsection{BDT analysis results}
\label{Sec_BDTCut}

\begin{figure}[tbp]
\centering 
	\includegraphics[width=0.8\linewidth]{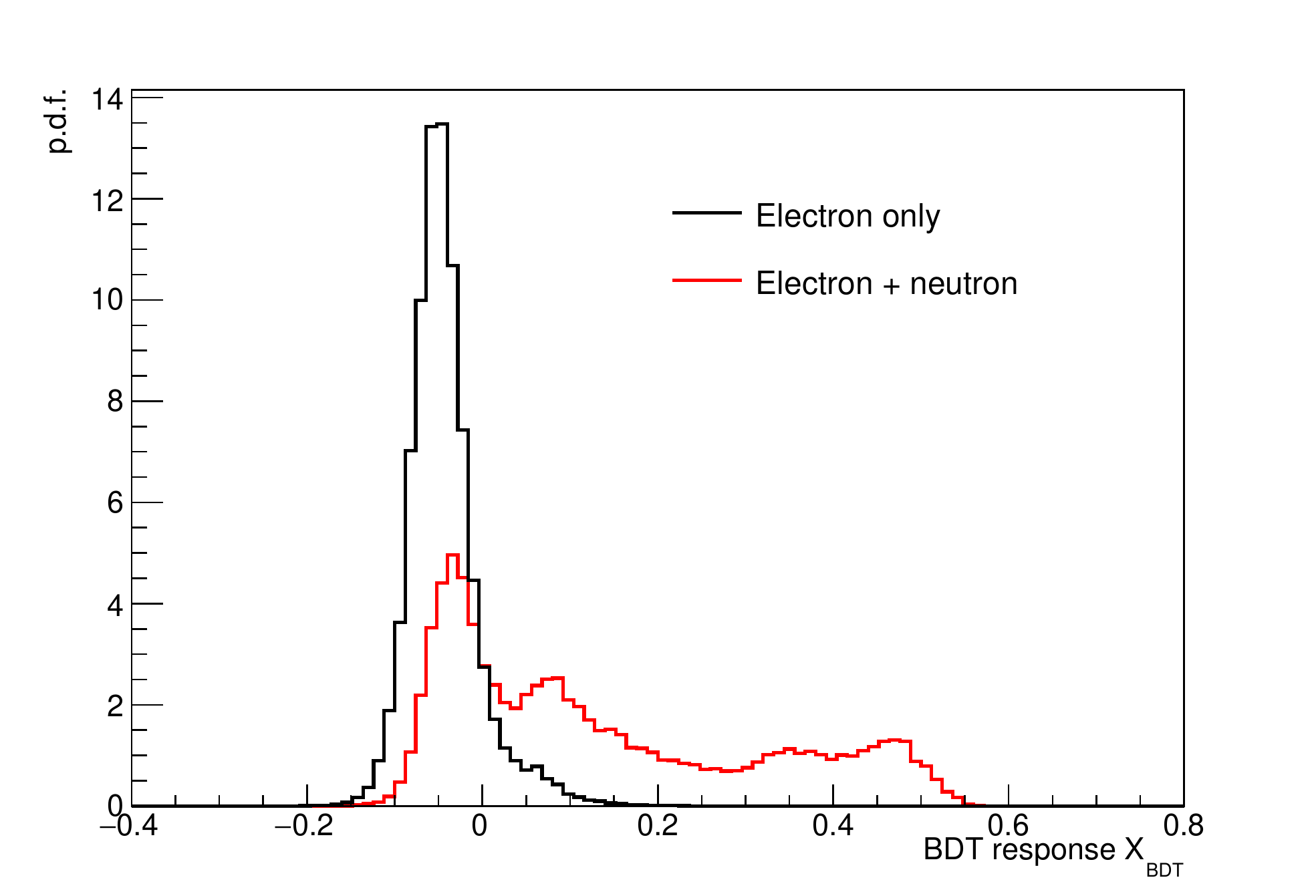}
	\caption{ $X_{BDT}$ value distribution for electrons only sample (black line) and for the electron and nuclear recoil sample (red line). }
	\label{fig:BDTdistribution}
\end{figure}

Figure~\ref{fig:BDTdistribution} shows the $X_{BDT}$ distribution from BDT analysis for each hypothesis ($H_0'$ and $H_1'$). 
Black and red lines correspond respectively to the $X_{BDT}$ distribution for the electron only sample (without $\mathrm{^7Li}$) and for the electron and nuclear recoil sample (with $\mathrm{^7Li}$).
The electron only sample presents a slightly asymmetric peak centered at -0.04 with its $X_\mathrm{BDT}$ value ranging from -0.2 to 0.2.
The nuclear recoil and electron sample present the same structure as observed previously for $X_\mathrm{BDT}$ value lower than 0, showing that our analysis classify some events as electron recoils in the sample (with $\mathrm{^7Li}$).

\begin{figure}[tbp]
\centering 
	\includegraphics[width=0.49\linewidth]{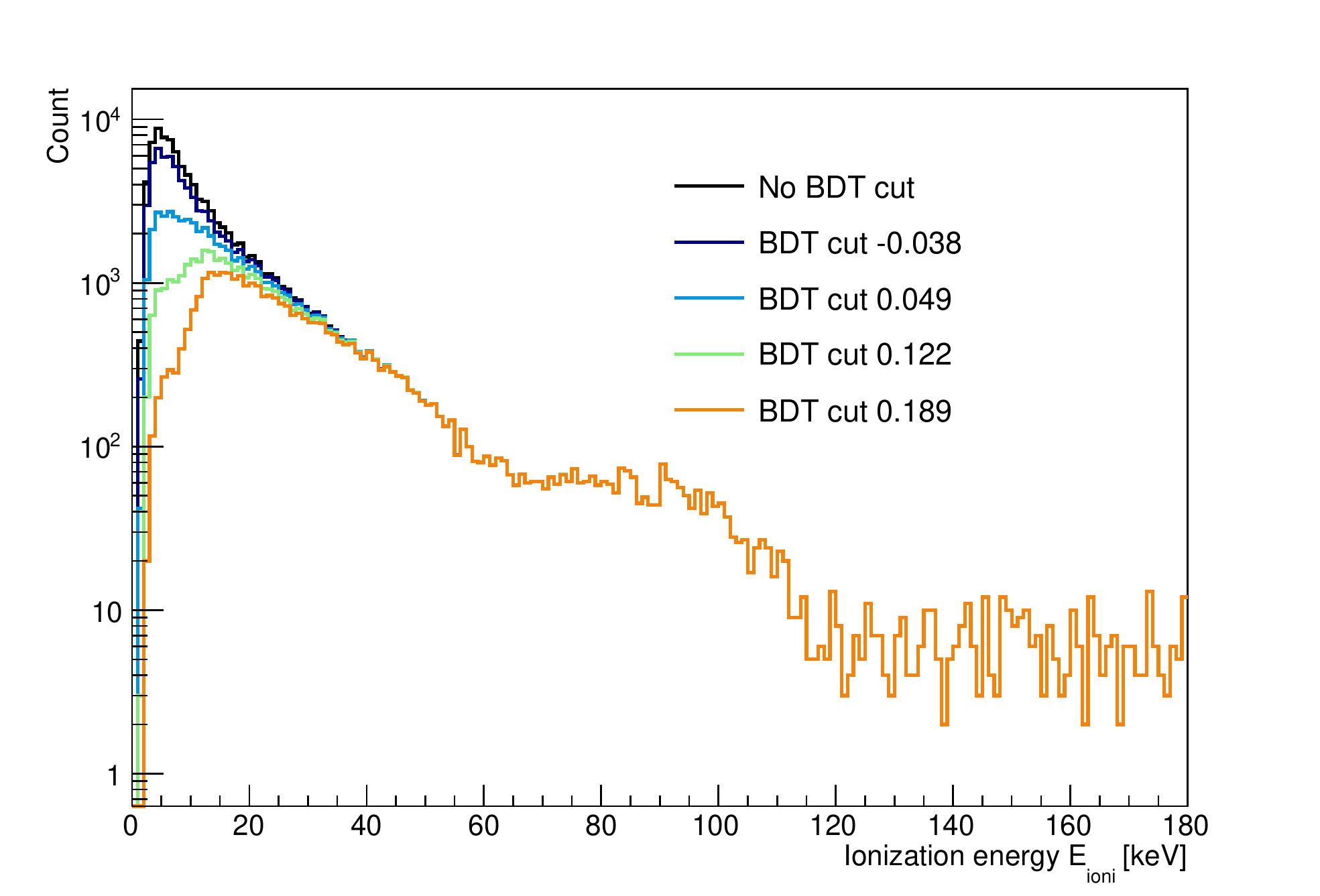}
	\includegraphics[width=0.49\linewidth]{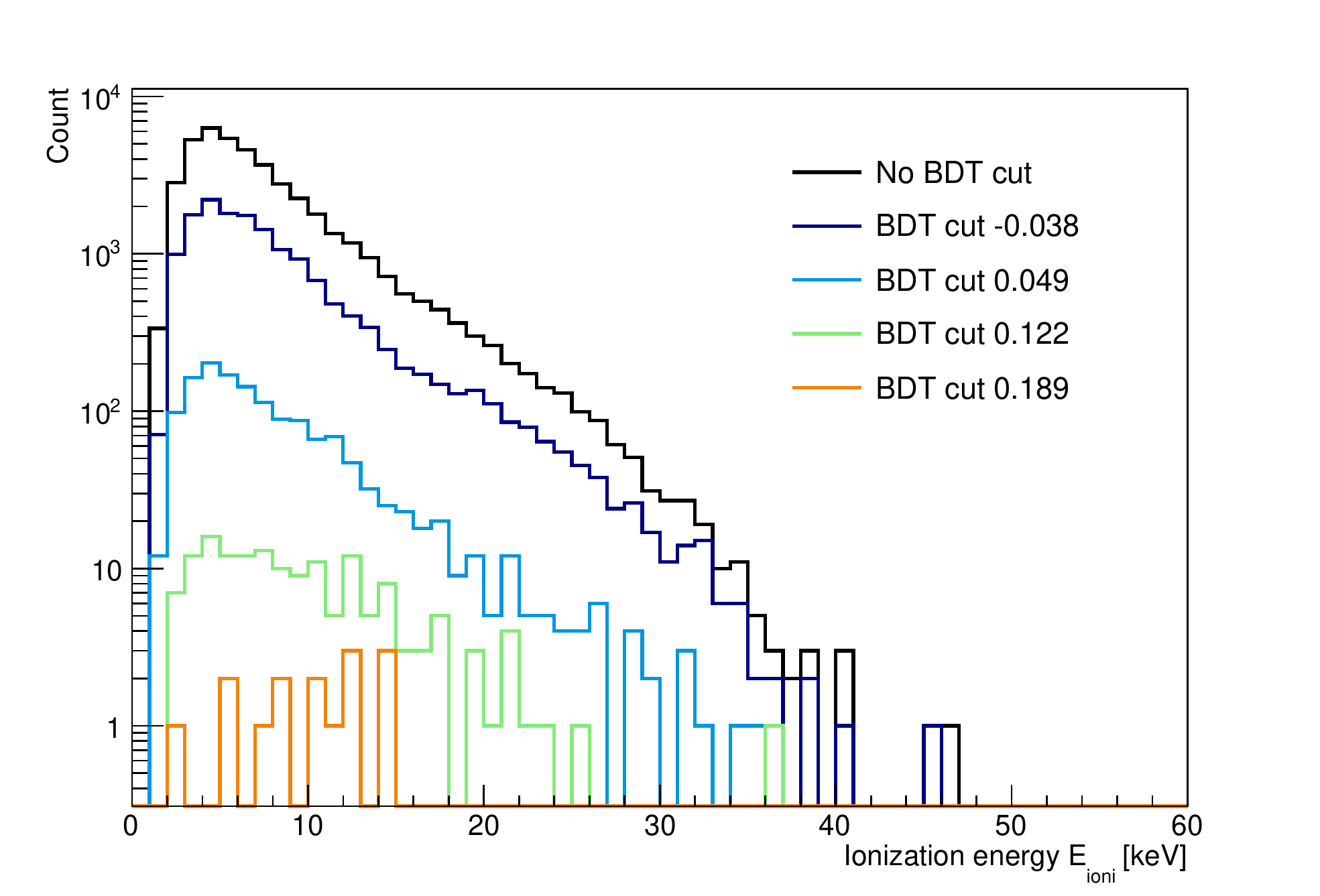}
	\caption{Energy spectra (left panel: with \Nucl{Li}{7} (electron and nuclear recoil), 
         	      right panel: without \Nucl{Li}{7} (electrons only) obtained after the application of cuts on the $X_{BDT}$ value as listed in table~\ref{tab:Cut}. Black lines represent energy spectra obtained without $X_{BDT}$ cut. }
	\label{fig:ResBDT}
\end{figure}

The rejection power of a cut on the $\mathrm{X_{BDT}}$ value is defined as the ratio between the total electron event number $N_{elec}$ and the number of selected electron event $N^{sel}_{elec}$:
\begin{equation}
\xi_R = \frac{N_{elec}}{N^{sel}_{elec}}
\end{equation}
This number corresponds to the size of the sample needed to have only one electron event passing the cut on the $\mathrm{X_{BDT}}$ value. It quantifies the goodness of the cut.
From the distribution of $f(X_\mathrm{BDT} | H_0)$ as shown by figure~\ref{fig:BDTdistribution} (black line), it is possible to determine the value of a cut on $\mathrm{X_{BDT}}$ associated with a certain rejection power. Table~\ref{tab:Cut} represents the value of the cut on $\mathrm{X_{BDT}}$ for rejection power ranging from $10^{2}$ to $10^{5}$.
The impact of these cuts on the experimental data is shown by figure~\ref{fig:ResBDT}. The left and right panels show respectively measured energy spectra with and without $\mathrm{^7Li}$ target after the application of table~\ref{tab:Cut} cuts. 
\renewcommand{\arraystretch}{1.2}
\begin{table}[tbp]
\centering
	\begin{center}
		\begin{tabular}{cc}
			\hline
			Rejection power $\xi_R$ & BDT cut$X_{BDT}^{cut}$ \\
			\hline
			\hline
			$10^{2}$ & $-0.038$\\
			$10^{3}$ & $0.049$\\
			$10^{4}$ & $0.122$\\
			$10^{5}$ & $0.188$\\
			\hline
		\end{tabular}
		\caption{Association of rejection power $\xi_R$ with cuts on $X_{BDT}$ value obtained from $X_{\mathrm{BDT}}$ value distribution for electron recoil sample presented in figure~\ref{fig:BDTdistribution}.}
		\label{tab:Cut}
	\end{center}
\end{table}
Right panel of figure~\ref{fig:ResBDT} shows that as $\mathrm{X_{BDT}}$ cut increases, the high energy contribution (above 15~keV) to the electron energy spectrum is reduced. This effect is also visible on the left panel of figure~\ref{fig:ResBDT}: we can observe that a low energy contribution (below 20~keV) to the energy spectrum coming from electron is suppressed.

In conclusion, we showed that crossing all the MIMAC observables in a BDT analysis, we are able to reach a $10^5$ rejection power level in the whole energy range.

\begin{figure*}[tbp]
\centering 
	\includegraphics[height=0.85\linewidth,angle=90,origin=c]{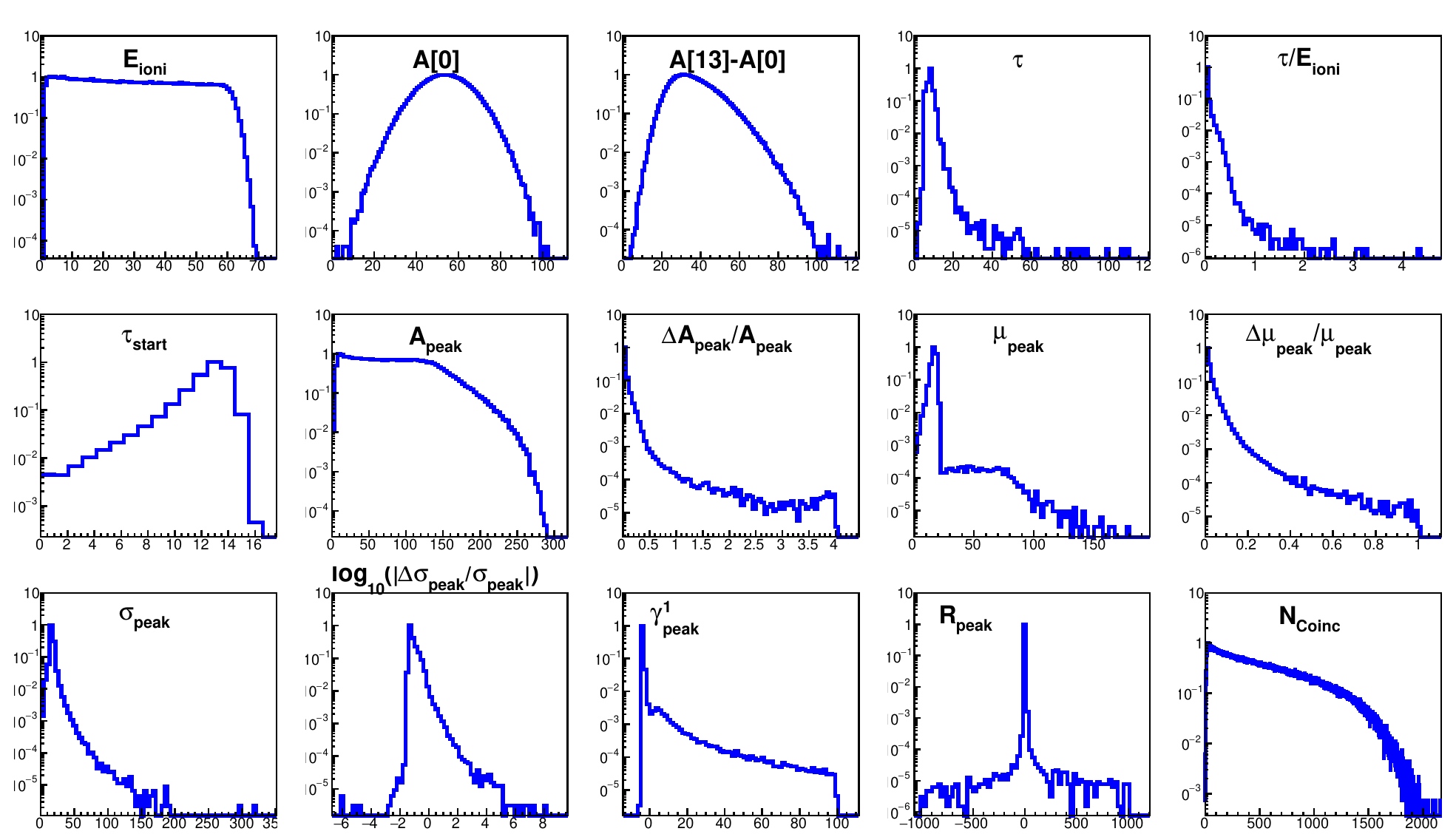}
	\caption{ One-dimension distribution of each discriminating observables obtained from Monte Carlo for fluorine nuclear recoils (part 1/2). Only events triggering the detector and passing minimal cuts are represented. }
	\label{fig:DiscriVar_MC_1}
\end{figure*}

\begin{figure*}[tbp]
\centering 
	\includegraphics[height=0.85\linewidth,angle=90,origin=c]{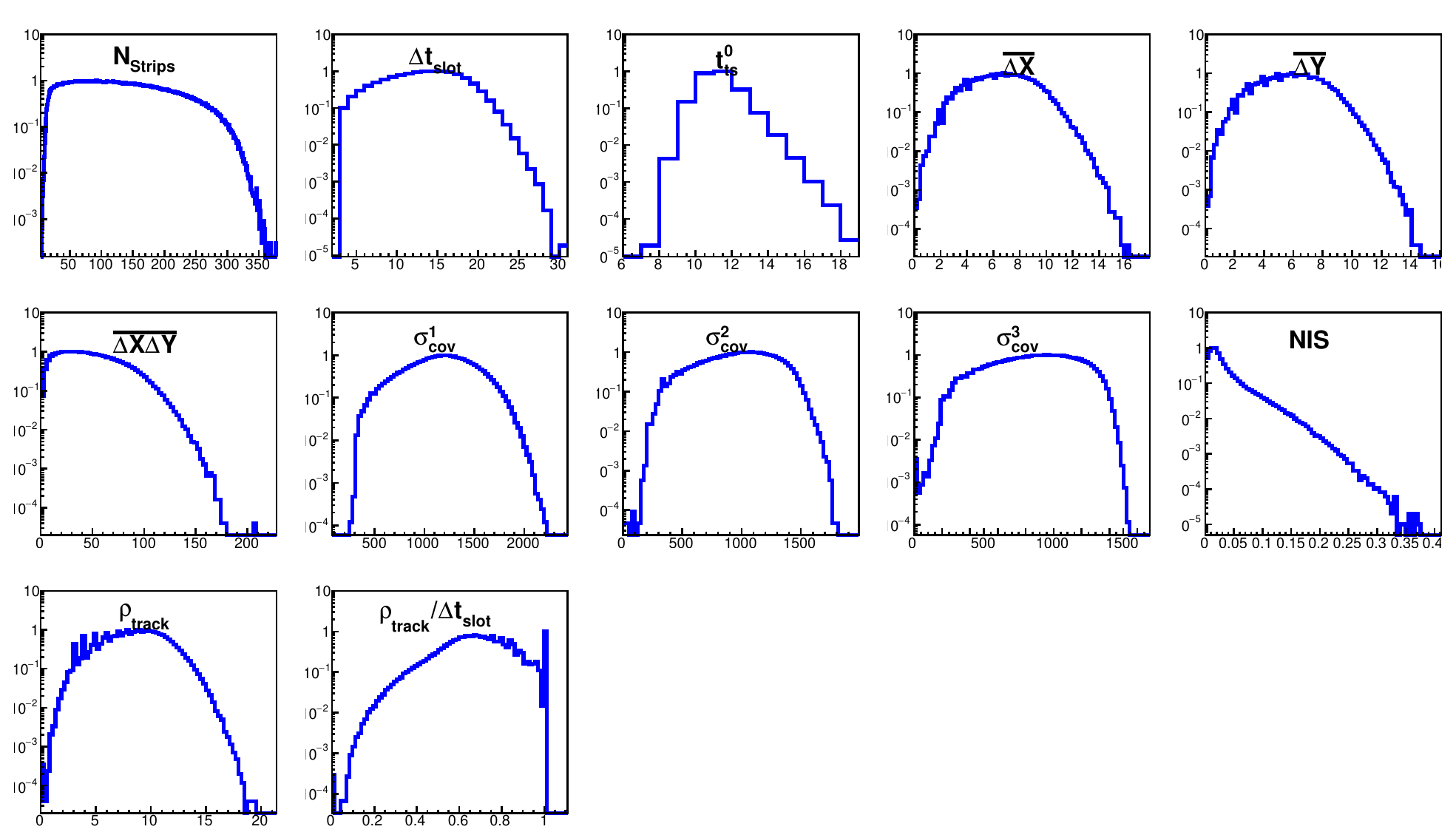}
	\caption{One-dimension distribution of each discriminating observables obtained from Monte Carlo for fluorine nuclear recoils (part 2/2). Only events triggering the detector and passing minimal cuts are represented. }
	\label{fig:DiscriVar_MC_2}
\end{figure*}

\section{BDT Analysis efficiency estimation}

For rare event searches, it is essential to estimate the analysis efficiency which takes place in WIMP-nucleus scattering event rate estimation.
It quantifies the probability for a nuclear recoil to be identified as a nuclear recoil.
In order to estimate this quantity, as the sample of nuclear recoils is a mixed sample containing electrons, we develop a Monte-Carlo simulation of fluorine nuclear recoil detection by the MIMAC detector.

 \subsection{Fluorine nuclear recoil simulation}
 \label{sec-Generation}
  
In order to estimate the primary ionization electron distribution along tracks of fluorine nuclear recoils we used the SRIM software.
$6.9\times 10^6$ fluorine nuclear recoils were simulated with energies ranging from 1~to~100~keV in the MIMAC gas mixture at 50~mbar. 
Each track was randomly distributed in $4\pi$ in the whole active volume to scan every possible direction of nuclear recoils in dark matter search data. 
Taking into account the electron drift velocity and the transverse and longitudinal diffusion coefficients, 
estimated by Magboltz~\cite{Biagi1999}, we estimated the ionization electron distribution in the anode plan as a function of time. 
Then, using the micromegas geometry and the flash-ADC transfer function, we were able to model MIMAC raw data. 
Finally, using the observable reconstruction software, we obtained the observable distribution for a set of fluorine nuclear recoils as presented in figures~\ref{fig:DiscriVar_MC_1}~and~\ref{fig:DiscriVar_MC_2}. We can see that our Monte-Carlo simulation is able to reproduce the different observables for fluorine recoils.

\subsection{BDT analysis efficiency estimation}

\begin{figure}[tbp]
\centering 
\begin{center}
		\includegraphics[width=0.8\linewidth]{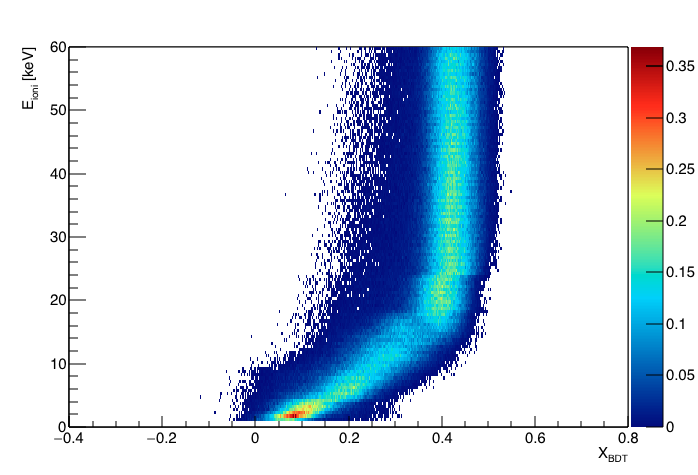}		
		\caption{Fluorine event probability density function on the plane $(X_{BDT},E_{ioni})$ estimated from Monte Carlo. The BDT classification previously obtained was applied to Monte-Carlo events generated as described in section~\ref{sec-Generation}.}
		\label{fig:BDTdistribSimu}
\end{center}
\end{figure}

The BDT classification, obtained previously, was applied on the simulated fluorine nuclear recoils.
Figure~\ref{fig:BDTdistribSimu} presents the probability density function of generated Monte-Carlo events in the plane $(X_{BDT},E_{ioni})$. We can note that the $X_{BDT}$ value increases as the ionization energy increases.

The efficiency of a BDT cut is defined as the ratio of the number of nuclear recoils passing the cuts $N^{sel}_{NR}$ and the total number of nuclear recoils $N_{NR}$: 
\begin{equation}
E = \frac{N^{sel}_{NR}}{N_{NR}}
\end{equation}
Table~\ref{tab:CutAcc} lists the analysis efficiency for each cut listed in table~\ref{tab:Cut} considering the full energy range and several thresholds. The uncertainty on efficiency is obtained by error propagation assuming that $N^{sel}_{NR}$ and $N_{NR}$ follow a Poisson statistics. The study of the impact of the systematics on the efficiency request a complete study and will be the subject of an ongoing publication.
We can note that a $10^{5}$ electron rejection power is reached with a $86.49 \pm 0.17\%$ nuclear recoil efficiency considering the full energy range and $94.67 \pm 0.19\%$ and $98.83 \pm 0.21\%$ if we assume a 5~keV and 10~keV thresholds respectively.

\begin{figure}[tbp]
\centering 
\begin{center}
		\includegraphics[width=0.8\linewidth]{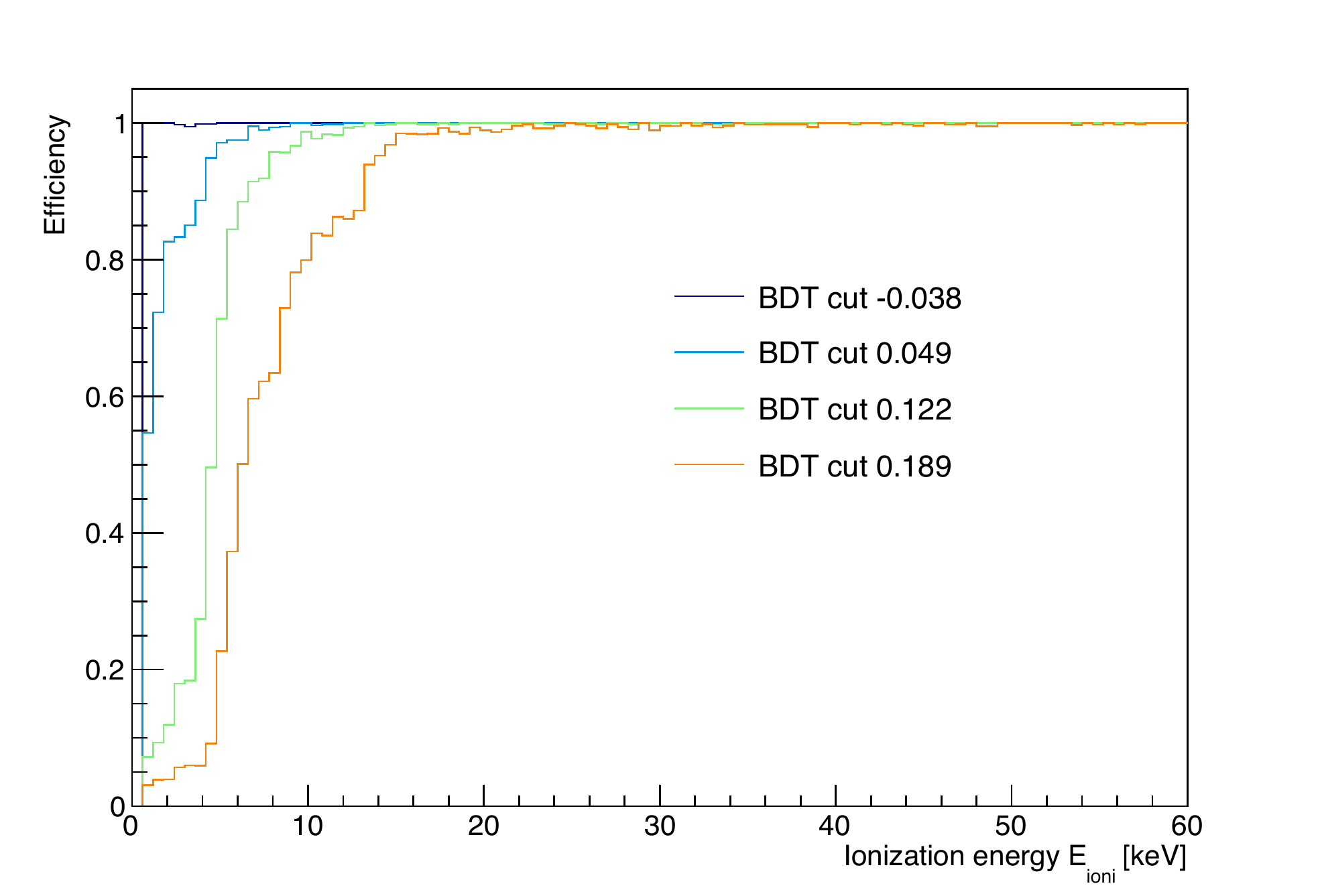}		
		\caption{ Analysis efficiency as a function of the ionization energy for several $X_{BDT}$ cut values as listed in table~\ref{tab:Cut}.}
		\label{fig:BDTSimu}
\end{center}
\end{figure}

Moreover, figure~\ref{fig:BDTSimu} represents the BDT analysis efficiency as a function of the ionization energy for several $X_{BDT}$ cuts listed in table~\ref{tab:Cut}. For each $X_{BDT}$ cut, we can see that the efficiency increases up to reach 100\% as the ionization energy increases.
In the case of a $10^5$ rejection power cut, we obtained a 50\% efficiency at 5 keV ionization energy with the present gain of the detector. This gain can be increased, if wished, to explore even better the low-energy range.

\begin{table}[tbp]
\centering
\begin{center}
\begin{tabular}{ccccc}
\hline
Rejection power & \multirow{2}*{$X_{BDT}$ cut} &Full range & \multicolumn{2}{c}{Efficiency with a lower threshold [\%] }\\
  $\xi_R$ & & efficiency [\%]& 5~keV & 10~keV \\
\hline
\hline
$10^{2}$ & $-0.038$ & $99.77 \pm 0.19$ & $99.88 \pm 0.20$ & $99.93 \pm 0.21$\\
$10^{3}$ & $0.049$ & $98.69 \pm 0.19$ & $99.74 \pm 0.20$ & $99.92 \pm 0.21$\\
$10^{4}$ & $0.122$ & $92.94 \pm 0.18$ & $98.70 \pm 0.20$ & $99.81 \pm 0.21$\\
$10^{5}$ & $0.188$ & $86.49 \pm 0.17$ & $94.67 \pm 0.19$ & $98.83 \pm 0.21$\\
\hline
\end{tabular}
\caption{Association of rejection power $\xi_R$ with cuts on $X_{BDT}$ value and the corresponding total simulated efficiency. The efficiency is obtained from Monte-Carlo model described in section~\ref{sec-Generation} and it is given for the full energy range and assuming a 5~keV and 10~keV thresholds. The given uncertainties only come from propagation of statistical uncertainty assuming a Poisson statistics.}
\label{tab:CutAcc}
\end{center}
\end{table}

\section{Conclusion}

In this paper, we proposed an original method for electron event rejection based on a multivariate analysis
applied to experimental data acquired using monochromatic neutron fields. 
This analysis shows that a $10^{5}$ rejection power is reachable for electron/recoil discrimination in the ionization energy range.
Moreover, the analysis efficiency was estimated showing that a $10^{5}$ electron rejection power is reached with a
$86.49 \pm 0.17\%$ nuclear recoil efficiency considering the full energy range and $94.67 \pm 0.19\%$ considering a 5~keV lower threshold.
The efficiency uncertainty does not take into account systematic uncertainties of the detector.

\section*{Acknowledgements}

We would like to thank M.~Pepino and A.~Martin for operating the AMANDE facility during the many experiments performed.
We acknowledge F.~Mayet for many helpful discussions.
The MIMAC collaboration acknowledges the ANR-07-BLAN-0255-03 funding.

\bibliography{Biblio}

\end{document}